\documentclass[aps,floatfix,showpacs,amsmath,amssymb]{revtex4}
\usepackage{natbib}
\usepackage{dcolumn}
\usepackage{graphicx}
\usepackage{color}
\usepackage[normalem]{ulem}
\newcommand{\be}{\begin{equation}}
\newcommand{\ee}{\end{equation}}
\newcommand{\bea}{\begin{eqnarray}}
\newcommand{\eea}{\end{eqnarray}}
\def\[{\begin{equation}}
\def\]{\end{equation}}
\begin{document}
%\tableofcontents
%
\title{The growth of structure in the Szekeres inhomogeneous cosmological models and the matter-dominated era}
\author{Mustapha Ishak\footnote{Electronic address: mishak@utdallas.edu}}
\author{Austin Peel}
\affiliation{
Department of Physics, The University of Texas at Dallas, Richardson, TX 75083, USA}
\date{\today}
\begin{abstract}
This study belongs to a series devoted to using the Szekeres inhomogeneous models in order to develop a theoretical framework where cosmological observations can be investigated with a wider range of possible interpretations. While our previous work addressed the question of cosmological distances versus redshift in these models, the current study is a start at looking into the growth rate of large-scale structure. The Szekeres models are exact solutions to Einstein's equations that were originally derived with no symmetries. We use here a formulation of the Szekeres models that is due to Goode and Wainwright, who considered the models as exact perturbations of a Friedmann-Lema\^itre-Robertson-Walker (FLRW) background. Using the Raychaudhuri equation we write, for the two classes of the models, exact growth equations in terms of the under/overdensity and measurable cosmological parameters. The new equations in the overdensity split into two informative  parts. The first part, while exact, is identical to the growth equation in the usual linearly perturbed FLRW models, while the second part constitutes exact non-linear perturbations. We integrate numerically the full exact growth rate equations for the flat and curved cases. We find that for the matter-dominated cosmic era, the Szekeres growth rate is up to a factor of three to five stronger than the usual linearly perturbed FLRW cases, reflecting the effect of exact Szekeres non-linear perturbations. We also find that the Szekeres growth rate with an Einstein-de Sitter background is stronger than that of the well-known non-linear spherical collapse model, and the difference between the two increases with time. This highlights the distinction when we use general inhomogeneous models where shear and a tidal gravitational field are present and contribute to the gravitational clustering. Additionally, it is worth observing that the enhancement of the growth found in the 
Szekeres models during the matter-dominated era could suggest a substitute to the argument that dark matter is needed when using FLRW models to explain the enhanced growth and resulting large-scale structures that we observe today.
\end{abstract} 
\maketitle
%
%\tableofcontents 
%
%
%%%%%%%%%%%%%%%%%%%%%%%%%%%%%%%%%%%%%%%%%%%%%%%%%%%%%%%%%%%%%%%%%%%%%%%%%%%
\section{Introduction}
%%%%%%%%%%%%%%%%%%%%%%%%%%%%%%%%%%%%%%%%%%%%%%%%%%%%%%%%%%%%%%%%%%%%%%%%%%%
%
Our modern era of cosmology has led to not only a wealth of astronomical
observations, but also to two major conundrums, namely, dark matter and dark
energy. It is therefore essential to explore theoretical frameworks that
allow for a wider range of possible interpretations of the cosmological data.

Such a framework can be provided by inhomogeneous cosmological models that
are exact solutions to Einstein's equations. A good deal of theoretical work
has been done on such models in the exact theory of {general relativity}, but
little work has been done to compare them to observations. 
This is not surprising, since it is not straightforward in these models to derive
observable functions ready to be compared to cosmological data. See, for example,
\cite{Celerier2007,krasinski1997,KSHM,HellabyAndKrasinski2002,SussmanAndTriginer1999,Bolejko2006a,Iguchietal2002,Alnesetal2006,Enqvist1,Garfinkle2006,Kaietal2006,Biswasetal2006,TanimotoNambu2007,Hunt,Enqvist,IshakEtAl2008,Chung,Garcia} and references therein.

This study is part of a series where we consider the Szekeres 
inhomogeneous models in order to develop a framework in which to analyze 
current and future cosmological observations. The models were originally 
derived by Szekeres in \cite{Szekeres1,Szekeres2} as an exact solution with 
a general metric that has no symmetries and has an irrotational dust source.  
The models are regarded as the best exact solution candidates to represent 
the true lumpy universe we live in. For example, the models are put in the 
same classification as the observed lumpy universe in \cite{EllisVanElst1998}.
The models have been investigated analytically and numerically by several {authors; see, for example,}  \cite{
Szafron1977,SussmanAndTriginer1999,BonnorAndTomimura1976,Bonnoretal1977,GoodeAndWainwright1982a,GoodeAndWainwright1982b,Bonnor1985,Barrow,KandH,HandK,Bolejko2007,Krasinski2008,IshakEtAl2008,BolejkoCelerier,Nolan2007,Nwankwo2010,PlebanskiAndKrasinski2006}. 

Any cosmological model must pass at least two types of cosmological
tests. The first is related to measurements of the expansion history and
cosmological distances, and the second is related to measurements of the
growth rate of structure formation. In previous studies
\cite{IshakEtAl2008,Nwankwo2010}, we explored distances versus redshift in the Szekeres models, while the current work looks into the growth of structure.

In this paper, we study the growth rate of structure in flat and curved cases of Class-I and Class-II Szekeres models using a reformulation of the models that was
introduced by Goode and Wainwright
\cite{GoodeAndWainwright1982a,GoodeAndWainwright1982b}, where the models can
be considered as non-linear exact perturbations of the
Friedmann-Lema\^itre-Robertson-Walker (FLRW) background. This formulation is
well-suited for the study of the growth and structure formation in these
models. The work of Ref. \cite{MeuresAndBruni} extended the flat case of Class-II in \cite{GoodeAndWainwright1982a} to include a cosmological constant and a discussion of the growth. Our work here extends that of \cite{GoodeAndWainwright1982a} to spatially curved cases in Class-I and Class-II in analyzing the growth. We also express the growth differential equations {as well as equations for the shear and tidal gravitational field,} all in terms of the under/overdensity and measurable cosmological parameters, {which offer} further insights {over the metric functions themselves}. We analyze the time evolution of the growth factor, {shear}, and tidal gravitational field scalars in flat and curved cases and discuss their interrelationship via the propagation equations to {produce stronger} gravitational clustering in Szekeres models.  

The paper is organized as follows. After presenting the formalism in section II, we show in section III how the Raychaudhuri propagation equation gives exact nonlinear growth equations in terms of the under/overdensity that divide into two meaningful parts. We express these equations for the flat and curved cases of Class-II in terms of growth factor, the scale factor, and measurable cosmological parameters. {We} perform numerical {integrations} of these new equations and plot the results for various cases. The results are also compared to those of the linearly perturbed Einstein-de Sitter model and the nonlinear spherical collapse. In section IV we repeat the analysis for models of Class-I. We then explore the shear and the tidal gravitational field and their relation to the growth in section V. We conclude in section VI. Throughout the paper we use units such that $8\pi G=c=1$.

%%%%%%%%%%%%%%%%%%%%%%%%%%%%%%%%%%%%%%%%%%%%%%%%%%%%%%%%%%%%%%%%%%%%%%%%%%%%%%%%%
\section{The Szekeres Models in the Goode and Wainwright representation}
%%%%%%%%%%%%%%%%%%%%%%%%%%%%%%%%%%%%%%%%%%%%%%%%%%%%%%%%%%%%%%%%%%%%%%%%%%%%%%%%%
%%%
Since the original derivation of the models by Szekeres in \cite{Szekeres1,Szekeres2}, they have been reformulated in at least two other different sets of coordinates. The second formulation was proposed by Goode and Wainwright \cite{GoodeAndWainwright1982a,GoodeAndWainwright1982b}, and a third set was used in e.g. \cite{HellabyAndKrasinski2002,PlebanskiAndKrasinski2006}. The relationships between these formulations can be found in \cite{GoodeAndWainwright1982b,krasinski1997,PlebanskiAndKrasinski2006}. As mentioned earlier, we chose in this study to use the representation of Goode and Wainwright because it is well-suited for the study of the growth of structure and exact perturbations of a smooth FLRW background. In order to be self-contained, we repeat and summarize here the presentation of the models in the Goode and Wainwright presentation (some typos have been fixed from their original paper). This also serves to set the notation to be used in the paper. The Szekeres metric in the Goode and Wainwright set of coordinates $\{t,x,y,z\}$ reads 
\be
ds^2= -dt^2+S^2\left[e^{2\nu}(dx^2+dy^2)+H^2W^2 dz^2\right],
\label{eq:metric}
\ee
where 
\bea
H(t,x,y,z)&=&A(x,y,z)-F(t,z)\nonumber \\
          &=&A(x,y,z)-(\beta_+f_++\beta_-f_-),
\label{eq:H}
\eea
and $H(t,x,y,z)$, $S(t,z)$, and $W(z)$ are all positive functions. $\beta_+$ and $\beta_-$ are functions of $z$ only, and $f_+$ and $f_-$ are functions of $t$ and $z$. $A(x,y,z)$ and $\nu(x,y,z)$ will be specified further below.

The source of the spacetime is irrotational dust, and the coordinates are comoving and synchronous. The cosmic dust fluid thus has the 4-velocity vector $u^a=[1,0,0,0]$.
 
The two classes of solutions are listed below with further specifications of the functions for each class. The function $S(t,z)$ satisfies the generalized Friedmann equation
\be
\dot{S}^2=-k+\frac{2M}{S},
\label{Friedmann}
\ee
where $\dot{}=\partial / \partial t$, $M=M(z)$, and $k=0,\pm 1$. Next, it can be verified that the functions $f_+$ and $f_-$ (see equations (\ref{fplus}) and (\ref{fminus}) below) are the increasing and decreasing solutions of the following ordinary differential equation \cite{GoodeAndWainwright1982b}  
\be
\ddot{F}+ 2\frac{\dot{S}}{S}\dot{F}-\frac{3M}{S^3}F=0,
\label{Ray1}
\ee
which can be derived from the field equations as well as from the Raychaudhuri equation as we discuss further in the following section.
  
Interestingly, the time evolution of the two spatial classes of the models is fully described by the same two equations (\ref{Friedmann}) and (\ref{Ray1}). Equation (\ref{Friedmann}) reduces to the usual Friedmann equation for a fixed $z$ while equation (\ref{Ray1}) governs density fluctuations.  

The matter density in the models is given by 
\be
\rho(t,x,y,z)=\frac{6M}{S^3}\left( 1+\frac{F}{H} \right) =\frac{6MA}{S^3H}.
\label{density}
\ee

In this formulation, the vanishing of the functions $\beta_{+}$ and $\beta_{-}$ is the necessary and sufficient condition for the models to reduce to FLRW models. 
In this limit, the sign of the matter density is determined by that of $M(z)$, and thus we shall limit our interest to models with $M(z)>0$.
%
%%%%%%%
\subsection{Time dependence}

The time dependence of the models is specified by the parametric and implicit solutions for the functions $S(t,z)$, $f_+(t,z)$, and $f_-(t,z)$ from equations (\ref{Friedmann}) and (\ref{Ray1}). The solution of the generalized Friedmann equation (\ref{Friedmann}) is given in parametric form using $\eta$ as:
\be
S=M \frac{dh(\eta)}{d\eta}\,\,\,\,{\rm{with}}\,\,\,\,t-T(z)=Mh(\eta),
\label{param}
\ee
where
\[
  h(\eta) = \left\{ 
  \begin{array}{l l}
    \eta-\sin\eta,  & \quad k=+1\\
    \sinh\eta-\eta, & \quad k=-1\\
    \eta^3/6,       & \quad k=0.\\
  \end{array} \right. \label{h}
\]
(Note that $M(z)>0$ is required in all cases, and that when $k=0$ or $k=-1$ then we require that $\dot{S}>0$.) So the scale function $S$ has the same time dependence as that of an FLRW dust model.
Next, the solution to equation (\ref{Ray1}) gives    
\[
  f_{+} = \left\{ 
  \begin{array}{l l}
    (6M/S)\,[1-(\eta/2)\cot(\eta/2)]-1, & \quad k=+1\\
    (6M/S)\,[1-(\eta/2)\coth(\eta/2)]+1, & \quad k=-1\\
    \eta^2/10, & \quad k=0\\
  \end{array} \right.\label{fplus}
\]
\[
  f_{-} = \left\{ 
  \begin{array}{l l}
    (6M/S)\,\cot(\eta/2), & \quad k=+1\\
    (6M/S)\,\coth(\eta/2), & \quad k=-1\\
    24/\eta^3, & \quad k=0.\\
  \end{array} \right.\label{fminus}
\]
%%%%
%%%%%%%%%%%%%
\subsection{Spatial dependence}
%%%%%%%%%%%%%
%%%%%
The spatial dependence of the models separates them into two classes.

{\begin{center}{Class-I: $S=S(t,z),\,S_z\ne0,\, f_{\pm}=f_{\pm}(t,z),\,T=T(z),\,M=M(z)$,}\end{center}}
with
\be
e^\nu=f(z)[a(z)(x^2+y^2)+2b(z)x+2c(z)y+d(z)]^{-1},
\ee 
\bea
ad&-&b^2-c^2=\epsilon/4,\,\quad \epsilon=0,\pm 1, \nonumber \\
A&=&f\nu_z-k\beta_+,\, \quad W^2=(\epsilon-kf^2)^{-1},\nonumber \\
\beta_+&=&-kfM_z/(3M),\, \quad \beta_-=fT_z/(6M),
\label{ClassIBeta}
\eea
where the subscript $z$ in the equations means differentiation with respect to z.

{\begin{center}{Class-II: $S=S(t),\,f_{\pm}=f_{\pm}(t),\,T=const,\,M=const$,}\end{center}}
with 
\be
e^\nu=[1+k/4\,(x^2+y^2)]^{-1},\,k=0,\pm1,\, \quad W=1,
\ee
\[
  A = \left\{ 
  \begin{array}{l l}
    e^\nu\{a(z)[1-\frac{k}{4}(x^2+y^2)]+b(z)x+c(z)y\}-k\beta_+,\, \quad k=\pm 1\\
    a(z)+b(z)x+c(z)y-\beta_+(x^2+y^2)/2,\,\,\,\,\,\,\quad k=0.\\
  \end{array} \right.
\]
\\
%%
%%%%%%%%%%%%%%%%%%%%%%%%%%%%%%%%%%%%%%%%%%%%%%%%%%%%%%%%%%%%%%%%%%%%%%%%%%%%%%%%%
\subsection{Raychaudhuri equation and gravitational attraction}
%%%%%%%%%%%%%%%%%%%%%%%%%%%%%%%%%%%%%%%%%%%%%%%%%%%%%%%%%%%%%%%%%%%%%%%%%%%%%%%%%
The complete dynamics of a cosmological model can be expressed in terms of a set of evolution and propagation equations (see, for example, \cite{Ellis1971,EllisVanElst1998} and citations therein). One of the fundamental propagation equations is the Raychaudhuri equation \cite{Raychaudhuri}, and it can be viewed as the basic equation of gravitational attraction \cite{EllisVanElst1998}. In the case of the Szekeres irrotational dust, the equation reads 
\be
\dot{\Theta}+\frac{1}{3}\Theta^2+2 \sigma^2+\frac{1}{2}\rho=0,
\label{Ray2}
\ee 
which includes the following quantities associated with the 4-velocity vector $u^a$:  
\begin{itemize}
\item{ {the} rate of volume expansion scalar 
\be
\Theta\equiv u^{a}{}_{;a}=3\frac{\dot{S}}{S}-\frac{\dot{F}}{H}
\label{expansion}
\ee}
\item{ {the} rate of shear tensor 
\be
\sigma_{a b}\equiv
 u_{(a; b)}+\dot{u}_{(a}u_{b)}-\frac{\Theta}{3}(g_{ab}+u_{a}u_{b}),
\ee 
where for our models the corresponding non-zero components are  
\be
2\sigma^x{ }_x=2\sigma^y{ }_y=-\sigma^z{ }_z=-\frac{2}{3}\frac{\dot{F}}{H}
\label{shear}
\ee
}     
\item{ {and the} matter density, $\rho$,  {which} is given by equation (\ref{density}).}
\end{itemize}
 {In the above, a semicolon denotes} covariant differentiation, and parentheses around indices indicate symmetrization.

For a complete Raychaudhuri equation including pressure, vorticity, 4-acceleration, and a cosmological constant (all zero in our models here), see for example \cite{Ellis1971,EllisVanElst1998}.

Now, using equations (\ref{expansion}), (\ref{shear}), the generalized Friedmann equation  {(\ref{Friedmann}) (making the appropriate substitution for $\ddot{S}$)}, and the density equation (\ref{density}), we can write the Raychaudhuri equation (\ref{Ray2}) as second-order ordinary differential equation  {that is} linear in the function $F$:
\be
\ddot{F}+ 2\frac{\dot{S}}{S}\dot{F}-\frac{3M}{S^3}F=0.
\label{Ray3}
\ee

As pointed out in \cite{GoodeAndWainwright1982b}, this differential equation of the metric function $F$ derives from the field equations as well as the Raychaudhuri equation. However, as we will explore in the rest of the paper, associating this equation with the meaning of the Raychaudhuri equation and that of other evolution equations involving shear and tidal gravitational fields will help us building a consistent discussion of gravitational clustering in the Szekeres models.

Including the previous section, all equations up to this point apply generally to both Class-I and Class-II of the Szekeres models. However in the following sections, we need to treat the discussion of the growth equations for Class-I and Class-II separately in view of their different spatial dependencies.
%%%%
%%%%%%%
%%%%%%%%%%%%%%%%%%%%%%%%%%%%%%%%%%%%%%%%%%%%%%%%%%%%%%%%%%%%%%%%%%%%%%%%%%%%%%%%%
\section{Structure growth exact equations in Szekeres Class-II models}
%%%%%%%%%%%%%%%%%%%%%%%%%%%%%%%%%%%%%%%%%%%%%%%%%%%%%%%%%%%%%%%%%%%%%%%%%%%%%%%%%
%%%
%%%%
%%
We start by exploring Class-II first because of its simpler spatial dependence. 
In this case, the function $S(t)$ is only a function of $t$ while $M$ and $T$ are constants. The connection to the growth of large scale structure is simpler since $\bar{\rho}$ in equations (\ref{delta}) and (\ref{rho}) below is only a function of $t$. 

Now we define, in the usual way, the  density contrast  $\delta$ as 
\be
\delta (t,x,y,z) \equiv \frac{\rho (t,x,y,z) - \bar{\rho}(t)}{\bar{\rho}(t)},
\label{delta}
\ee
so that we can write 
\be
\rho(t,x,y,z)=\bar{\rho}(t)[1+\delta(t,x,y,z)].
\label{rho}
\ee
Comparing equation (\ref{rho}) and equation (\ref{density}) specialized to Class-II, that is 
\be
\bar{\rho}(t)=\frac{6M}{S(t)^3}
\ee
which is a function of $t$ only since $M$ is constant, we identify immediately 
\be
\delta(t,x,y,x)= \frac{F}{H}.
\label{identification}
\ee

Using the relations $\dot{F}=\dot{\delta}H+\delta\dot{H}$ and $\ddot{F}=\ddot{\delta}H+2\dot{\delta}\dot{H}+\delta\ddot{H}$ obtained from equation (\ref{identification}), the Raychaudhuri equation (\ref{Ray3}) becomes
\be
\ddot{\delta}H+2\dot{\delta}\dot{H}+\delta\ddot{H}+2\frac{\dot{S}}{S}(\dot{\delta}H+\delta\dot{H})-3\frac{M}{S^3}\delta H = 0.
\ee
Noting that $\dot{H}=-\dot{F}$ and $\ddot{H}=-\ddot{F}$, we can rewrite the above equation (divided by $H$) as
\be
\ddot{\delta}+2\frac{\dot{H}}{H}\dot{\delta}-3\frac{M}{S^3}\delta^2+2\frac{\dot{S}}{S}\dot{\delta}-3\frac{M}{S^3}\delta = 0.
\ee
We can eliminate $\dot{H}/H$ with the above relation for $\dot{F}$ and arrive at the following exact growth equation for the Szekeres Class-II models:
\be
\ddot{\delta}+2\frac{\dot{S}(t)}{S(t)}\dot{\delta}-3\frac{M}{S(t)^3}\delta-\frac{2}{1+\delta}{\dot{\delta}}^2-3\frac{M}{S(t)^3}\delta^2=0.
\label{growth1}
\ee

Now, we use the same idea of Goode and Wainwright \cite{GoodeAndWainwright1982a,GoodeAndWainwright1982b}, who stated that the Szekeres models in their formulation allows one to compare them with linear perturbations of the FLRW models \cite{GoodeAndWainwright1982a}. Additionally, as explained in \cite{GoodeAndWainwright1982b}, the function $S(t,z)$ in the generalized Friedmann equation (\ref{Friedmann}) corresponds to the scale factor in the FLRW models in the sense that for each value of $z$ it satisfies the Friedmann equation. In other words, for each Szekeres model explored, we associate a corresponding linearly perturbed FLRW model. The exact growth equation obtained from the Szekeres model is then compared to the linearly perturbed associated FLRW model. The unperturbed FLRW model is what is called here and in other papers the associated FLRW background. It is worth clarifying that this association is not based on the usual procedure where limits are imposed on some coordinates or special values imposed on the metric functions such that the Szekeres models will reduce to an FLRW model \cite{krasinski1997}, even if it may be related to it.  

So following \cite{GoodeAndWainwright1982a,GoodeAndWainwright1982b}, we associate the Szekeres models to non-linear exact perturbations of an associated FLRW background, and we write here the growth equation (\ref{growth1}) with an exact density contrast as perturbations of an FLRW model with the corresponding Hubble function and matter density as given below in equations (\ref{Hubble}) and (\ref{rho_FLRW}). The  equation then reads  
\be
\ddot{\delta}+2 \mathbb{H}(t)_{_{FLRW-B}}\dot{\delta}-4 \pi G \rho(t)_{_{FLRW-B}}\delta-\frac{2}{1+\delta}{\dot{\delta}}^2-4 \pi G \rho(t)_{_{FLRW-B}}\delta^2=0,
\label{growth2}
\ee
where we have identified the Hubble expansion rate of the FLRW background (FLRW-B):
\be
\mathbb{H}(t)_{_{FLRW-B}}\equiv \frac{\dot{S}(t)}{S(t)}.
\label{Hubble}
\ee
Note that this $\mathbb{H}(t)$ is not the same function as the metric function $H(t,x,y,z)$. Since $M$ is a constant in Class-II here, we have used equation (\ref{density}) to express the factor $\frac{3M}{S^3}$ in terms of the smooth background matter density as
\be
\frac{3M}{S(t)^3}=4\pi G \rho(t)_{_{FLRW-B}}.
\label{rho_FLRW}
\ee
 We note that while  {throughout} this treatment we have set $\kappa=8\pi G=1$, we restored its value here just to make the identification with the usual FLRW expressions immediate. 

It is also worth noting that the Raychaudhuri propagation equation provided us with an exact equation for the growth function $\delta$ that can be remarkably split into two meaningful parts.  {One part, which consists of the first three terms, is} identical to the usual equation of the growth for a linearly perturbed FLRW model.  {The other part contains the remaining nonlinear terms} in $\delta$ and $\dot{\delta}$ and is similar to second order perturbation terms. But we note here that our equation is exact, and $\delta$ does not have to be small.   
In the next sub-sections, we specialize the exact growth rate equation (\ref{growth1}) of the Szekeres Class-II to the flat and curved cases and then integrate them numerically. 
%%%%%%%%%%%%%%%%%%%%%%%%%%%%%%%%%%%%%%%%%%%
\subsection{Growth rate equations and integration in flat Szekeres Class-II models}
%%%%%%%%%%%%%%%%%%%%%%%%%%%%%%%%%%%%%%%%%%%
%%%%%%%%%%%
In this case, the corresponding FLRW background is thus the well-known Einstein-de Sitter model that is appropriate  for describing the matter-dominated phase of cosmic evolution. For an Einstein-de Sitter background, the scale factor can be solved from equation (\ref{Friedmann}) as 
\be
S(t)=\left(\frac{9}{2}M\right)^{1/3}t^{2/3}=t^{2/3},
\label{scaleF}
\ee
where for $k=0$ we can set $M=2/9$ in all generality (for example, see \cite{GoodeAndWainwright1982b}). Using equation (\ref{scaleF})  {in} equation (\ref{growth2}) or (\ref{growth1}) yields
\be
\ddot{\delta}+\frac{4}{3t}\dot{\delta}-\frac{2}{3t^2}\delta-\frac{2}{1+\delta}\dot{\delta}^2-\frac{2}{3t^2}{\delta}^2=0.
\label{growth4}
\ee

We note again that the first 3 terms are exactly the same as the ones in the usual linearly perturbed Einstein-de Sitter case, so this could be identified with the linear theory when $\delta$ is small. The second part is made of two nonlinear terms in $\dot{\delta}$ and $\delta$, and can be compared to second order perturbation terms \cite{Peebles1980}. 

The full equation obtained can be compared as well to the spherical nonlinear collapse model, and we do that further below. As in the case of nonlinear spherical collapse \cite{Peebles1980} for Einstein-de Sitter, a parametric solution is well-known (see, for example, \cite{GaztanagaLobo}), but we found it more practical for our cosmology numerical codes to perform the integrations numerically. Moreover, the numerical integration schemes are expandable to cases where parametric solutions are not known.   

Next, we write the growth rate equation (\ref{growth4}) in terms of the scale factor. To convert time derivatives to $S$ derivatives, we first note that
\be
\dot{\delta} = \frac{2}{3S^{1/2}}\delta', 
\ee
and
\be
\ddot{\delta} = \frac{4}{9S}\delta' - \frac{2}{9S^2}\delta'',
\ee
where primes denote derivatives with respect to $S$.  Here and throughout the rest of the paper, we write $a(t)$ for the scale factor $S(t)$ by analogy with FLRW models and take it to have the standard normalization $a(t_0)=1$ today. We thus obtain
\be
{\delta}''+\frac{3}{2}\frac{{\delta}'}{a}-\frac{3}{2}\frac{{\delta}}{a^2} -\frac{2}{1+\delta}{\delta'}^2-\frac{3}{2}\frac{{\delta}^2}{a^2}=0.
\label{growth5}
\ee
For our numerical integrations, we write the growth equation in terms of the growth rate $G\equiv\delta/a$. Using the relations $\delta' = G + aG'$ and $\delta'' = aG'' + 2G'$, we get
\be
G''+\frac{7}{2}\frac{G'}{a}-\frac{2}{a}\frac{(G+aG')^2}{1+aG}-\frac{3}{2}\frac{G^2}{a}=0.
\label{growth6}
\ee

We integrate this equation using a fourth-order Runge-Kutta algorithm with adaptive step size \cite{Press}. We implement the Runge-Kutta code with the function vectors  
$ \textbf{y}=\{G,G'\}$ and $\frac{d\textbf{y}}{da}=\{G',G''\}$ 
\cite{Press}   so that the second-order ODE (\ref{growth6}) is reduced to two first-order ODEs  {that are immediately integrable}. 

\begin{figure}
\begin{center}
\begin{tabular}{|c|c|c|}
\hline
{\includegraphics[width=2.2in,height=2.25in,angle=-90]{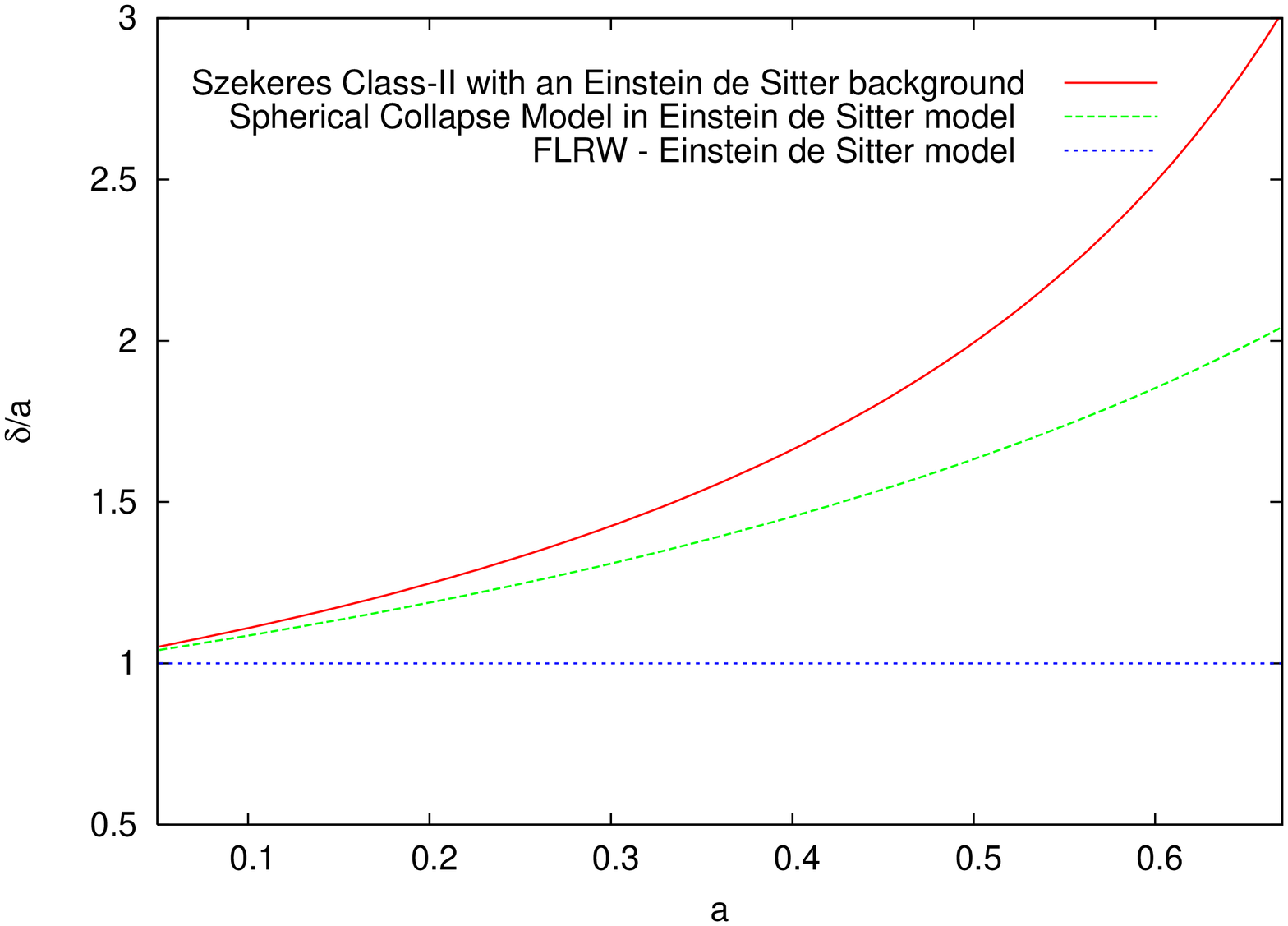}}&
{\includegraphics[width=2.2in,height=2.25in,angle=-90]{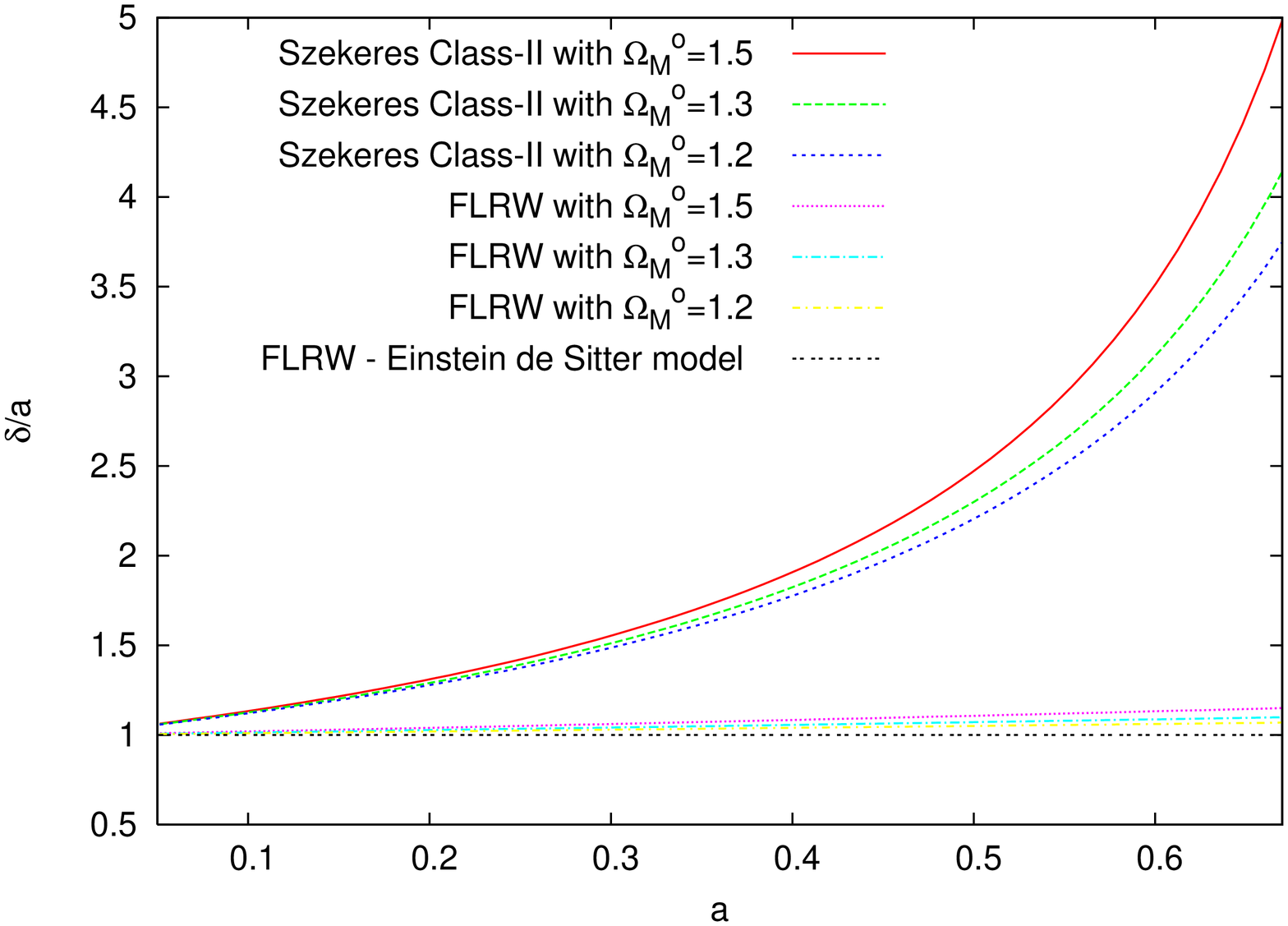}}&
{\includegraphics[width=2.2in,height=2.25in,angle=-90]{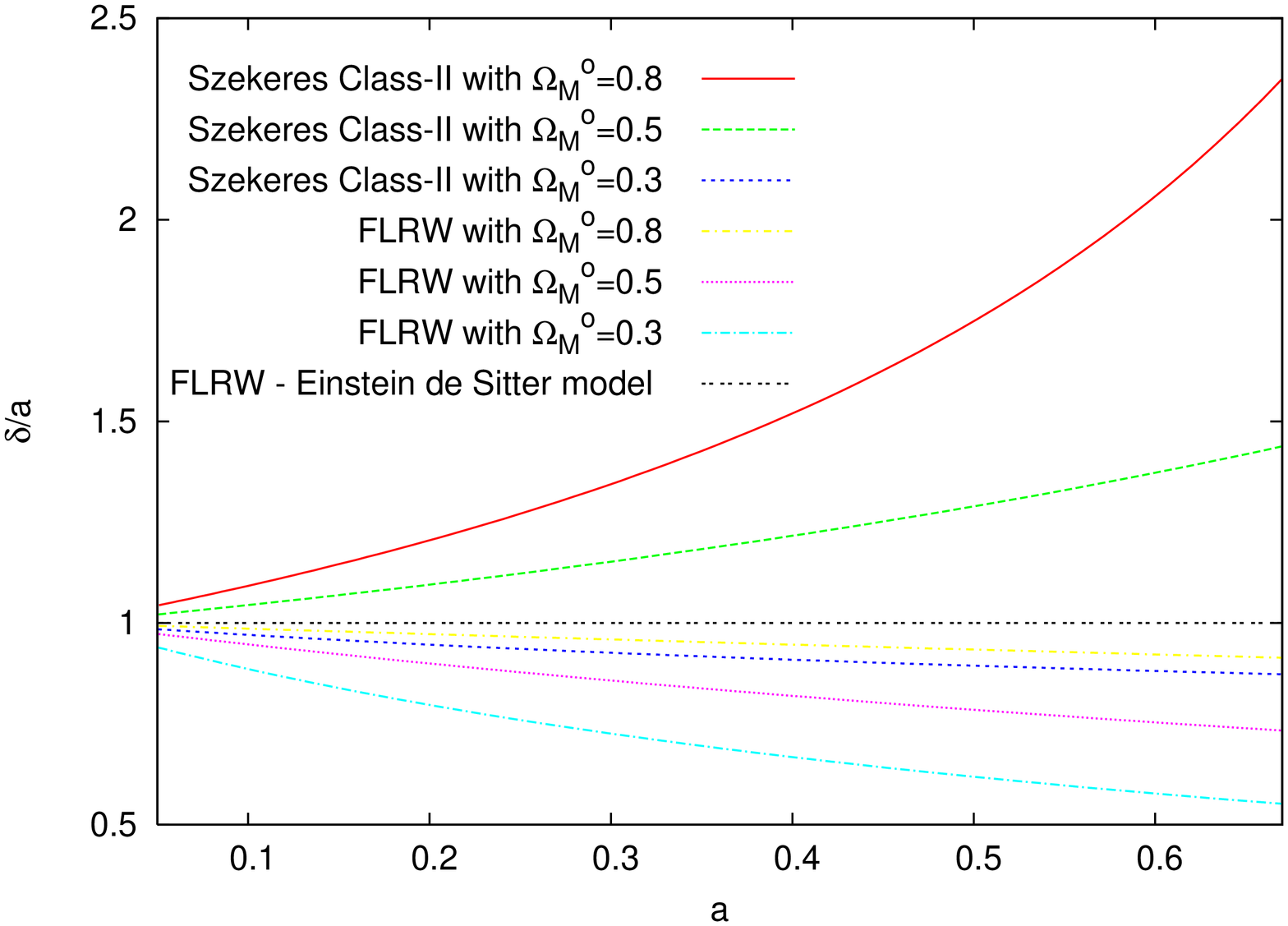}} \\
\hline  
\end{tabular}
\caption{\label{fig:Growth}
LEFT: Growth rate of structure in flat Szekeres Class-II models (or Class-I models with a fixed value of $z$; see section IV) (solid-red curve), the usual spherical collapse model (green-dashed), and the perturbed Einstein-de Sitter (EdS) model (blue-dotted). The Szekeres growth rate is stronger than that of the perturbed EdS by up to a factor of 3. The Szekeres growth rate is also stronger than that of the spherical collapse model.
CENTER: Growth rate of structure in positively curved Szekeres Class-II models for various values of $\Omega_M^0$ (or Class-I models with a fixed value of $z$ and various values of $\Omega_M^0(z)$). The growth rate in linearly perturbed FLRW models with the same values of $\Omega_M^0$ are plotted for comparison. 
RIGHT: Growth rate of structure in negatively curved Szekeres Class-II models for various values of $\Omega_M^0$ (or Class-I models with a fixed value of $z$ and various values of $\Omega_M^0(z)$). The growth rate in linearly perturbed FLRW models with the same values of $\Omega_M^0$ are plotted for comparison as well. 
In both cases, the Szekeres growth rates are stronger than those of the corresponding perturbed FLRW models by up to 5 times. 
} 
\end{center}
\end{figure}

Our results for the integration are given in the left part of figure \ref{fig:Growth}. The horizontal line $G=1$ corresponds to the linearly perturbed Einstein-de Sitter model, while the red curve (highest) represents the solution to the full exact growth equation (\ref{growth6}) for the flat Szekeres model of Class-II. The Szekeres growth is up to a factor of 3 stronger than that of the perturbed Einstein-de Sitter background. 

For further comparisons, we also integrate numerically the growth for the well-known nonlinear spherical collapse model in an Einstein-de Sitter background  {(see, for example, \cite{Peebles1980})}, where the governing equation is given by  
\be
{\delta}''+\frac{3}{2}\frac{{\delta}'}{a}-\frac{3}{2}\frac{{\delta}}{a^2} -\frac{4}{3}\frac{1}{1+\delta}{\delta'}^2-\frac{3}{2}\frac{{\delta}^2}{a^2}=0,
\label{spherical1}
\ee
or in the $G$-notation
\be
G''+\frac{7}{2}\frac{G'}{a}-\frac{4}{3a}\frac{(G+aG')^2}{1+aG}-\frac{3}{2}\frac{G^2}{a}=0.
\label{growth7}
\ee
The resulting integration is also plotted in the left part of figure 1. The Szekeres growth is found to be stronger than that of the spherical collapse model, as well. Moreover, the relative difference also increases as a function of time. This indicates that the growth rate is different when the spherical symmetry approximation for inhomogeneities is not assumed. As we discuss in section V further below and in the concluding remarks, the Szekeres models have shear and tidal gravitational fields that contribute to the enhancement of its gravitational collapse and growth rate.   
%%%
%%%%%%%%%%%%%%%%%%%%%%%%%%%%%%%%%%%%%%%%%%%%%%%%
\subsection{Growth rate equations and integration in curved Szekeres Class-II models}
%%%%%%%%%%%%%%%%%%%%%%%%%%%%%%%%%%%%%%%%%%%%%%%%
%
We derive now the growth ODE for the positively and negatively curved cases ($k=\pm1$). Both situations can be handled simultaneously and lead ultimately to the same equation that we integrate numerically. We begin from the growth equation (\ref{growth1})
\be
\ddot{\delta} + 2\frac{\dot{a}}{a}\dot{\delta} - 3\frac{M}{a^3}\delta - \frac{2}{1+\delta}\dot{\delta}^2 - 3\frac{M}{a^3}\delta^2 = 0,
\label{app1}
\ee
and the generalized Friedmann equation (\ref{Friedmann})
\be
\left( \frac{\dot{a}}{a} \right)^2 \equiv \mathbb{H}^2 = \frac{2M}{a^3} - \frac{k}{a^2},
\label{app2}
\ee
where an overdot denotes differentiation with respect to $t$ (partial in the case of $\delta$ and total for $a$). We define $\Omega_M \equiv 2M/(a^3\mathbb{H}^2)$ and $\Omega_k \equiv -k/(a^2\mathbb{H}^2)$ by analogy with standard cosmology such that (\ref{app2}) can be written as $\Omega_M + \Omega_k = 1$. We proceed with a more general approach than for the flat case that does not require an explicit expression for $a$ as a function of $t$ for the different $k$ values. This is accomplished by using equation (\ref{app2}) directly to recast equation (\ref{app1}) into integrable form. As before, we convert time derivatives to $a$ derivatives via the (now general) relations
\be
\dot{\delta} = \frac{\partial\delta}{\partial t} = \frac{\partial\delta}{\partial a}\frac{\mathrm{d}a}{\mathrm{d}t} = \delta'\dot{a}
\label{app3}
\ee
and
\be
\ddot{\delta} = \frac{\partial}{\partial t}(\delta'\dot{a}) = \dot{a}^2\delta''+\ddot{a}\delta'.
\label{app4}
\ee
Substituting these relations into equation (\ref{app1}), we find
\be
\delta''+\left(\frac{\ddot{a}}{\dot{a}^2}+\frac{2}{a}\right)\delta' - 3\frac{M}{a^3\dot{a}^2}\delta - \frac{2}{1+\delta}\delta'^2 - 3\frac{M}{a^3\dot{a}^2}\delta^2 = 0.
\label{app5}
\ee
We next use equation (\ref{app2}) to eliminate $\dot{a}$ and $\ddot{a}$ in favor of $\Omega_M$ and (implicitly) $\Omega_k$. Taking its time derivative and dividing by $\dot{a}^2$ gives the coefficient on $\delta'$:
\be
\frac{\ddot{a}}{\dot{a}^2}+\frac{2}{a} = \frac{4-\Omega_M}{2a}.
\label{app6}
\ee
The coefficients on $\delta$ and $\delta^2$ are also easily converted as
\be
3\frac{M}{a^3\dot{a}^2}=\frac{3}{2}\frac{\Omega_M}{a^2},
\label{app7}
\ee
and so we obtain the equation
\be
\delta'' + \left( 2 - \frac{\Omega_M(a)}{2} \right)\frac{\delta'}{a}-\frac{3}{2}\Omega_M(a)\frac{\delta}{a^2} -
	\frac{2}{1+\delta}\delta'^2 - \frac{3}{2}\Omega_M(a)\frac{\delta^2}{a^2} = 0.
\label{app8}
\ee
Again, we see that the first part agrees perfectly with the usual growth equation for the spatially curved linearly perturbed FLRW models \cite{Peebles1980}, while the second part is made of two nonlinear terms that can be compared to second order perturbation terms. Converting as before to $G$-notation for integration, we obtain
\be
G'' + \left( 4-\frac{\Omega_M(a)}{2} \right)\frac{G'}{a} + 2(1-\Omega_M(a))\frac{G}{a^2} - \frac{2}{a}\frac{(G+aG')^2}{1+aG}
	- \frac{3}{2}\Omega_M(a)\frac{G^2}{a} = 0.
\label{app9}
\ee
We must now express $\Omega_M$ explicitly in terms of the scale factor, $a$, and the matter density parameter today, $\Omega_M^0$. Using the definitions of $\Omega_M$ and $\Omega_k$ and the Friedmann equation (\ref{app2}) evaluated today, we see that
\be
1-\Omega_M^0 = \frac{-k}{\mathbb{H}_0^2}
\label{app10}
\ee 
allows the Friedmann equation (\ref{app2}) at a given time to be written as
\be
\mathbb{H}^2 = \frac{2M}{a^3} - \frac{\mathbb{H}_0^2(\Omega_M^0-1)}{a^2},
\label{app11}
\ee
where a super- or subscript naught means that the parameter is evaluated today. Dividing by $\mathbb{H}^2$, (\ref{app11}) becomes
\be
1 = \Omega_M(a) - \frac{\mathbb{H}_0^2(\Omega_M^0-1)}{\mathbb{H}^2a^2},
\label{app12}
\ee
and dividing (\ref{app2}) by $\mathbb{H}_0^2$ yields
\be
\frac{\mathbb{H}^2}{\mathbb{H}_0^2} = \frac{1}{a^3}[\Omega_M^0-a(\Omega_M^0-1)].
\label{app13}
\ee
Thus we find the usual relation
\be
\Omega_M(a) = \frac{\Omega_M^0}{\Omega_M^0+a(1-\Omega_M^0)},
\label{app14}
\ee
which, upon substituting into (\ref{app9}), gives finally
\begin{figure}
\begin{center}
\begin{tabular}{|c|c|c|}
\hline
{\includegraphics[width=2.25in,height=2.25in,angle=-0]{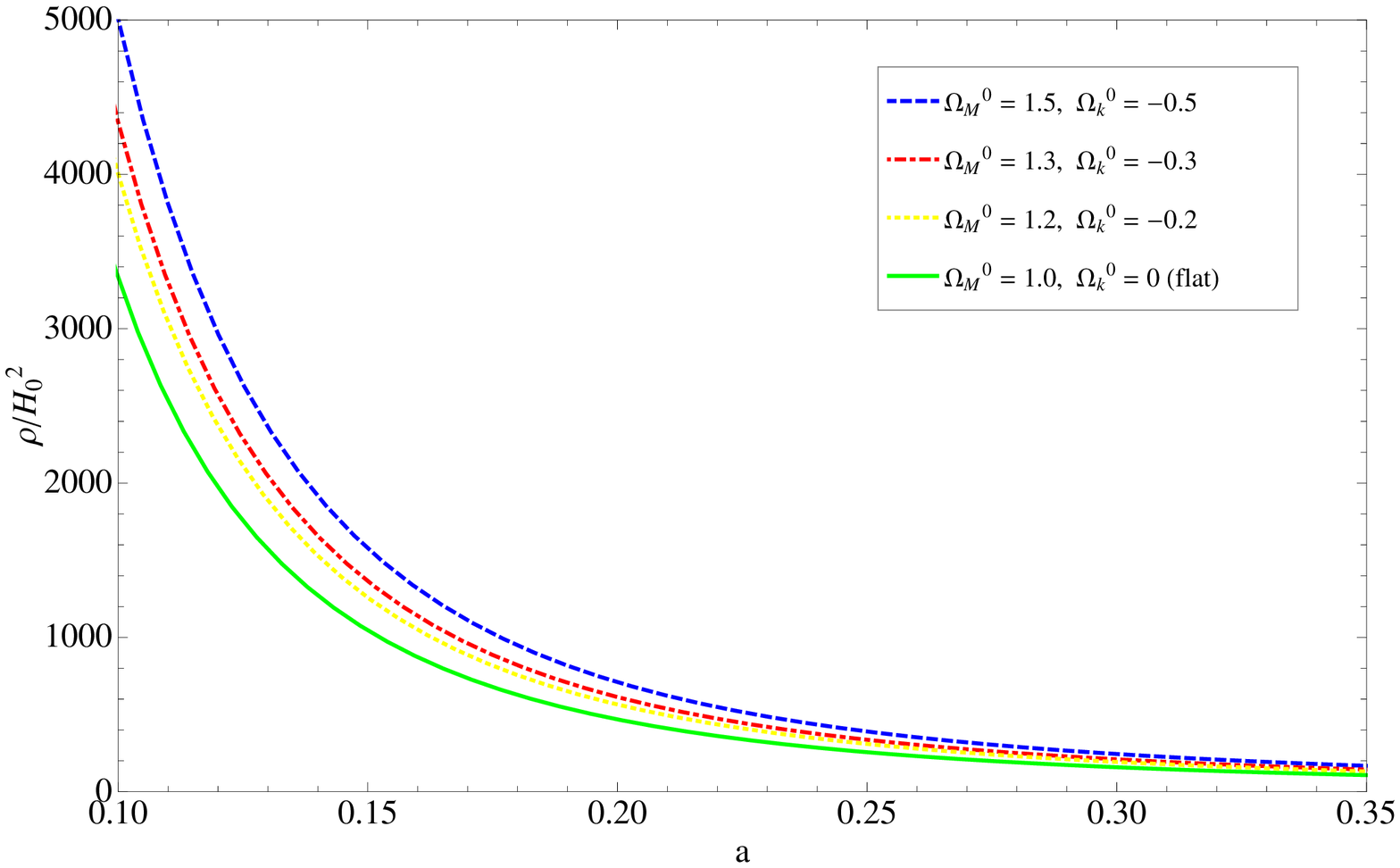}}&
{\includegraphics[width=2.25in,height=2.25in,angle=-0]{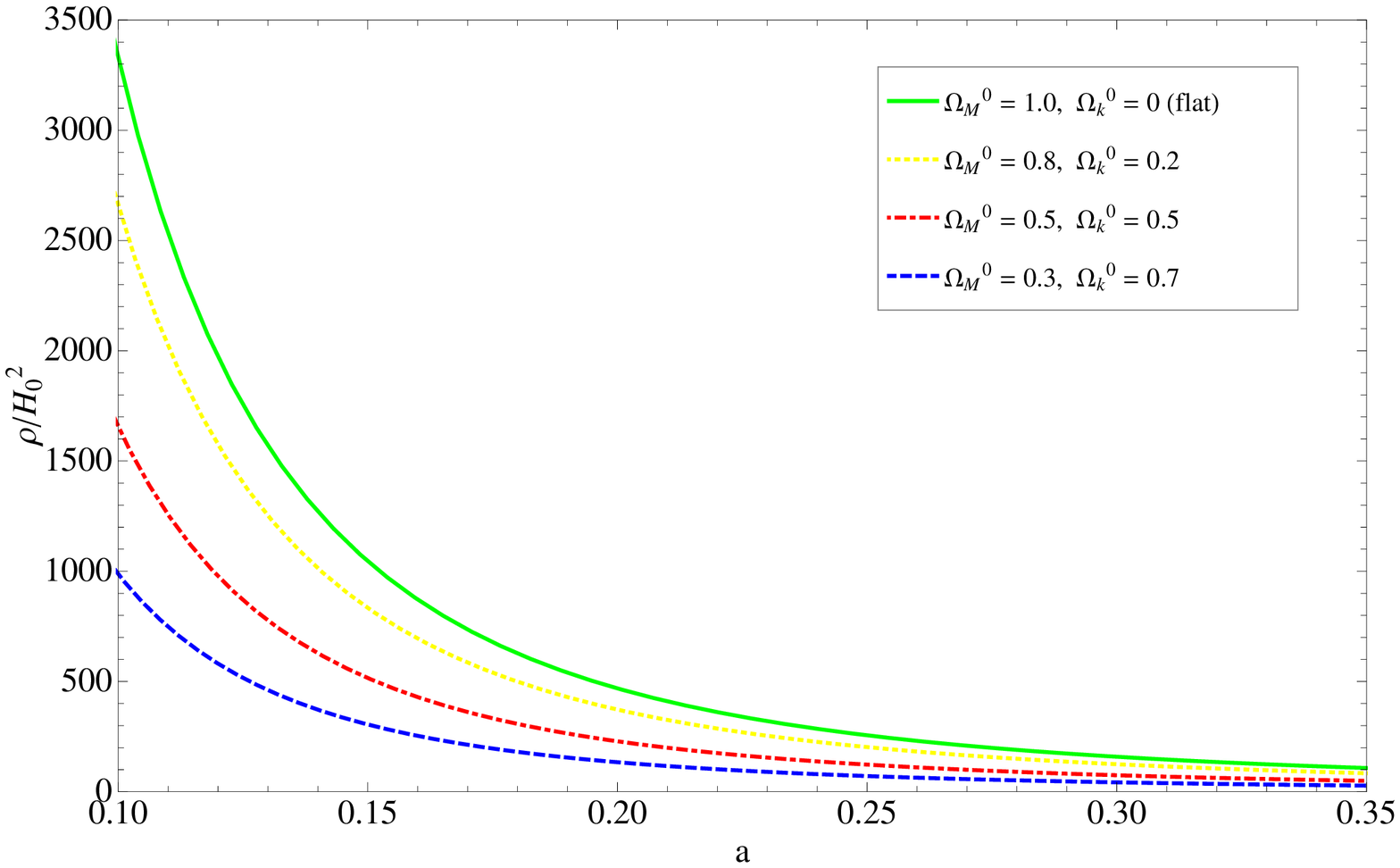}}&
{\includegraphics[width=2.25in,height=2.25in,angle=-0]{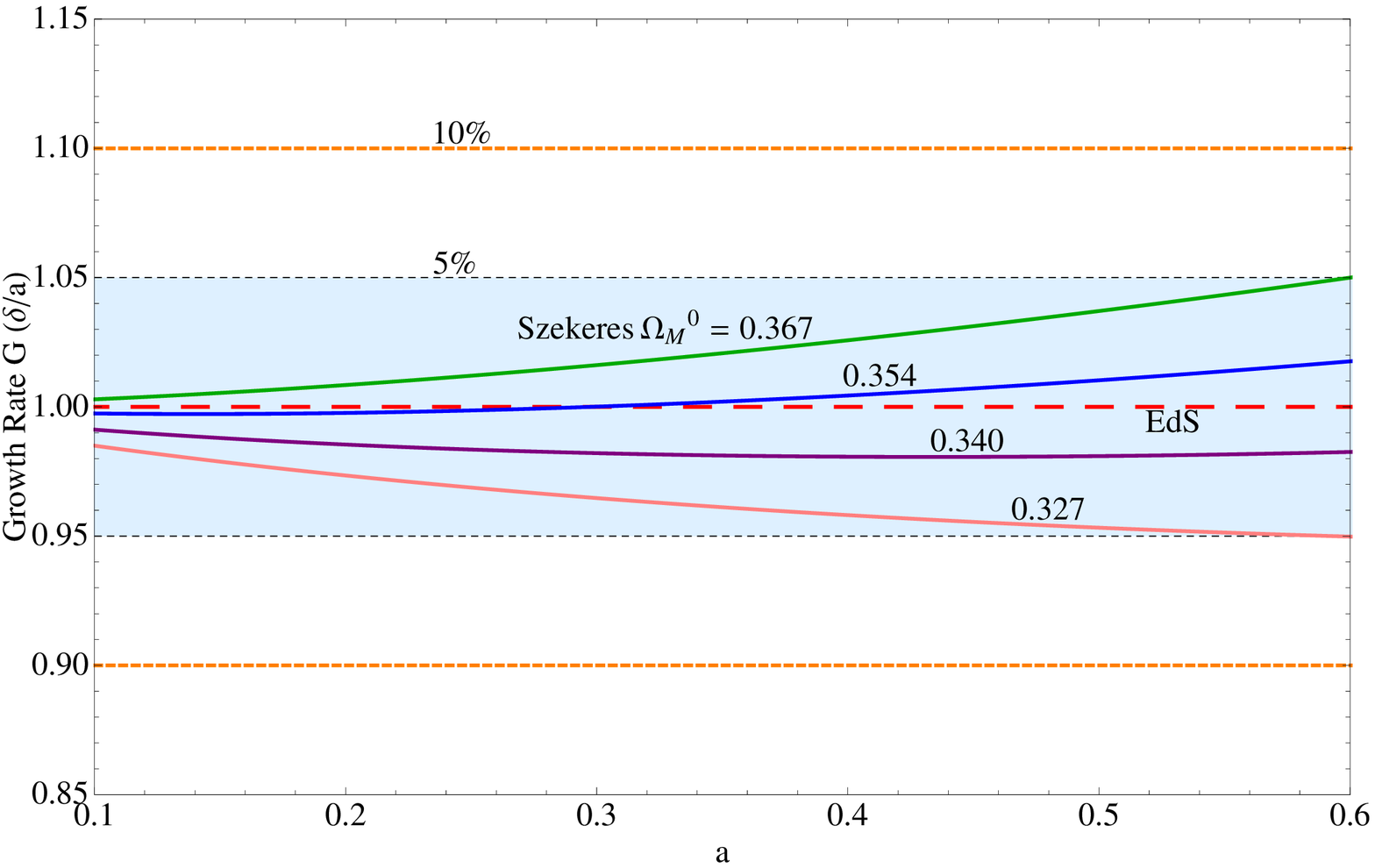}} \\
\hline  
\end{tabular}
\caption{\label{fig:density}
LEFT: Energy density in flat and positively curved Szekeres Class-II models (or Class-I models with a fixed value of $z$; see section IV). The density is plotted as a function of the scale factor {well within the matter-dominated} cosmological era ($a\le 0.35$) (i.e., well before $a\approx 0.60${, the scale factor associated with matter and cosmological constant equality} in the FLRW-LCDM standard model \cite{IshakReport,Carroll}). The density diverges as it should as $a$ {tends} to the initial singularity, {and it} decreases monotonically for increasing $a$ with no further divergences during the matter-dominated era. 
CENTER: Energy density in negatively curved Szekeres Class-II models (or Class-I models with a fixed value of $z$). The flat case is also given as reference. The same observations hold here. Further discussion is given in sections III-B and V. 
RIGHT: Comparisons of some Szekeres models to a growth range band of $5\%$ around the Einstein-de-Sitter growth, which is also the approximate representation for the FLRW-LCDM model during the matter dominated era under consideration. 
We find that there are Szekeres models that are consistent with such a growth range during the cosmological era considered, but the Szekeres models require only about a third of the matter density compared to that of Einstein-de Sitter.
This is consistent with our result of stronger structure growth in the Szekeres models.
} 
\end{center}
\end{figure}
\be
G'' + \left(4-\frac{\Omega_M^0}{2[\Omega_M^0+a(1-\Omega_M^0)]} \right)\frac{G'}{a}+2\left( 1-\frac{\Omega_M^0}{\Omega_M^0+a(1-\Omega_M^0)}\right)\frac{G}{a^2}
	-\frac{2}{a}\frac{(G+aG')^2}{1+aG} - \frac{3}{2}\left(\frac{\Omega_M^0}{\Omega_M^0+a(1-\Omega_M^0)}\right)\frac{G^2}{a} = 0.
\label{app15}
\ee
It is this equation that we integrate numerically in our code for various values of $\Omega_M^0$.
We note that the same differential equation (\ref{app15}) governs both the $k=+1$ and the $k=-1$ cases. The dependence on $\Omega_k$ is accounted for in the sign of $1-\Omega_M^0$. As expected, it is clear that setting $\Omega_M^0=1$ indeed recovers the growth equation for the $k=0$ case derived in the above sub-section.

Our results for the integration of the curved cases are given in figure 2 for various values of $\Omega_M^0$ and include positively and negatively curved models. We also integrate and plot the corresponding linearly perturbed FLRW models. As in the spatially flat cases, we find that the curved Szekeres growth is up to 5 times stronger than that of the linearly perturbed curved FLRW. Also, some Szekeres models with $\Omega_M^0<1$ can still have a stronger growth than that of the Einstein-de Sitter growth. 

In addition to the growth, it is worth plotting the energy density evolution in order to verify that, after the expected initial singularity, it has no other divergences during the {matter-dominated era of interest here}. We use equations (\ref{density}) and (\ref{rho}) and some of the steps above to write the energy density as follows:
\be
\rho=\frac{6M}{a^3}(1+\delta)=3\mathbb{H}_0^2 \frac{\Omega_M^0}{a^3} (1+aG).
\label{RhoToPlot}
\ee
Our plots in figure \ref{fig:density} for the density as a function of the scale factor for the flat and curved Szekeres show that the energy density diverges toward the initial singularity{, as it should, and} after it decreases monotonically with no other divergences during the {matter-dominated era} plotted here with $a\le 0.35$ (i.e., well before $a\approx 0.60${, which} corresponds to the equality time between matter dominance and cosmological constant dominance in an FLRW-LCDM (Lambda-Cold-Dark-Matter) model; see, for example, \cite{IshakReport,Carroll}). Our first goal here is to represent the growth rate during the matter dominated era with {$a\le 0.35$,} and as the density plots show, we are far away from any possible {pancake singularity that could occur at later times \cite{GoodeAndWainwright1982b}}. Additionally, such pancake {singularities} can be avoided in some cases by the addition of a cosmological constant to the models \cite{MeuresAndBruni}, although the discussion there was limited to the flat case.
Finally, we conclude in this section that the growth rate of large-scale structure is consistently found to be much stronger in the Szekeres models than in the linearly perturbed FLRW models with the same matter density.
%%
%%%%
%%%%%%%
%%%%%%%%%%%%%%%%%%%%%%%%%%%%%%%%%%%%%%%%%%%%%%%%%%%%%%%%%%%%%%%%%%%%%%%%%%%%%%%%%
\section{Structure growth exact equations in Szekeres Class-I models}
%%%%%%%%%%%%%%%%%%%%%%%%%%%%%%%%%%%%%%%%%%%%%%%%%%%%%%%%%%%%%%%%%%%%%%%%%%%%%%%%%
%%%
%%%%
\subsection{Flat Szekeres Class-I models}
%%%%%%%%%%%%%%%%%%%%%%%%%%%%%%%%%%%%%%%%%%%
%%%%%%%%%%%
%%
Unlike in the spatially curved Class-I models, the time evolution equations (\ref{param}) in the flat case can be decoupled and one can set without loss of generality $M=2/9$ \cite{GoodeAndWainwright1982b}. As a result, the treatment of the growth in the flat case can be developed in exactly the same way as was done for Class-II in section III-A. However, a closer look at equation (\ref{ClassIBeta}) shows that for $k=0$ in Class-I, $\beta_{+}=0$, so there are no growing modes and this case becomes of no interest for our purpose.  
%%%
%%%%%
%%%%%%%%%%%%%%%%%%%%%%%%%%%%%%%%%%%%%%%%%%%%%%%%
\subsection{Growth rate equations and integration in curved Szekeres Class-I models}
%%%%%%%%%%%%%%%%%%%%%%%%%%%%%%%%%%%%%%%%%%%%%%%%
%%%%%%%%%
In the spatially curved Class-I models, the functions $S(t,z)$ and $M(z)$ are now $z$-dependent. As a result, the identification of an under- or overdensity function as well as a background will require some discussions and definitions different from the ones we made for Class-II.  

We define a density contrast  $\hat\delta$ as 
\be
\hat\delta (t,x,y,z)\equiv \frac{\rho(t,x,y,z)\,\, - \rho_{q}(t,z)}{{\rho}_{q}(t,z)},
\label{delta_hat}
\ee
where we use the quasi-local average density \cite{Sussman1,SussmanAIP,Sussman2,SussmanBolejko}:
\be\
\rho_q(t,z)= \frac{\int_z\int_{x}\int_y \mathcal{F} \rho(t,x,y,z)\sqrt{-h}\, dz\, dx\, dy}{\int_z\int_x\int_y \mathcal{F} \sqrt{-h}\,dz\, dx \,dy}=\langle \rho \rangle _{q\,\mathcal{D}[z]}(t),
\label{quasi-local-density}
\ee
for the domain $\mathcal{D}[z]$ delimited by $z=$constant and contained in the hypersurface $t$=constant. Here $h$ is the determinant of the 3-dimensional part of the projection tensor $h_{ab}=g_{ab} +u_a u_b$, and $\mathcal{F}$ is a physically meaningful weighting function that was discussed in \cite{Sussman1,SussmanAIP,Sussman2,SussmanBolejko,Note}.

It was shown in \cite{Sussman1,SussmanAIP,Sussman2,SussmanBolejko} that $\rho_q$ is a coordinate independent quantity that can be expressed in terms of curvature invariants and is given for the Szekeres Class-I models by \cite{SussmanBolejko} 
\be
\rho_q(t,z)=\frac{6M(z)}{S(t,z)^3}.
\label{Szekeres_Rho_q}
\ee
As discussed in \cite{Sussman1,SussmanAIP,Sussman2}, the term quasi-local follows from the integral definition used for the quasi-local Misner-Sharp mass-energy in spherical symmetry \cite{Misner}. Bearing in mind that $M(z)$ is associated with such a quasi-local mass-energy, the result (\ref{Szekeres_Rho_q}) is not unexpected.  

Using the definition (\ref{quasi-local-density}) we can write  
\be
\rho(t,x,y,z)= \rho_q(t,z)[1+\hat\delta(t,x,y,z)].
\label{rho_2}
\ee
Comparing equation (\ref{rho_2}) and equation (\ref{density}), we identify
\be
\hat\delta(t,x,y,z) =\frac{F}{H}.
\label{identification2}
\ee
Again, using equation (\ref{identification2})  {in} the Raychaudhuri equation (\ref{Ray3}) gives,  after a few steps, the following exact growth equation for the Szekeres Class-I models:
\be
\ddot{\hat\delta}+2\frac{\dot{S}(t,z)}{S(t,z)}\dot{\hat\delta}-3\frac{M(z)}{{S(t,z)^3}}\hat\delta-\frac{2}{1+\hat\delta}{{\dot{\hat\delta}^2}}-3\frac{M(z)}{S(t,z)^3}\hat\delta^2=0.
\label{growth1_classI}
\ee
Next, following a large number of papers using the Lema\^itre-Tolman-Bondi models (LTB) for the purpose of comparing them to observations (see for example \cite{Enqvist,BAW,Quartin,Celerier2007} and references therein), we can generalize some definitions from the FLRW models to the Szekeres models.  We start by rewriting the Szekeres generalized Friedmann equation (\ref{Friedmann}) as 
\be
\left(\frac{\dot{S}(t,z)}{S(t,z)} \right)^2 \equiv \mathbb{H}(t,z)^2 = \frac{2M(z)}{S(t,z)^3} - \frac{k}{S(t,z)^2}.
\label{app2_ClassI}
\ee
We then define 
\be
\Omega_M (t,z) \equiv \frac{2M(z)}{S(t,z)^3\mathbb{H}(t,z)^2} 
\ee
and
\be 
\Omega_k(t,z) \equiv \frac{-k}{S(t,z)^2\mathbb{H}(t,z)^2}
\ee 
by analogy with standard cosmology such that (\ref{app2_ClassI}) can be written as 
\be
\Omega_M (t,z)+ \Omega_k(t,z) = 1.
\ee

Now, following the same approach we used for Class-II earlier and the idea introduced by Goode and Wainwright \cite{GoodeAndWainwright1982a,GoodeAndWainwright1982b}, we can define a background that is independent of coordinates $x$ and $y$ and where the $\hat\delta(t,x,y,z)$ can be regarded as an exact perturbation satisfying the exact equation 
\be
\ddot{\hat\delta}+2 \mathbb{H}(t,z)\dot{\hat\delta}-4 \pi G \rho_{q}\,\hat\delta-\frac{2}{1+\hat\delta}{\dot{\hat\delta}}^2-4 \pi G {\rho_q}\,  \hat\delta^2=0,
\label{growth2_ClassI}
\ee

Again, we can observe that this exact growth equation splits into two meaningful parts. One part, which consists of the first three terms, is similar to the usual equation of the growth for a linearly perturbed FLRW model. The other part consists of the nonlinear terms in $\hat\delta$ and $\dot{\hat\delta}$. 

We follow now the steps performed in the curved cases for Class-II with an overdot here always denoting partial differentiation with respect to $t$. We use equation (\ref{app2_ClassI}) to recast equation (\ref{growth1_classI}) into integrable form. After some steps similar to equations (\ref{app3})-(\ref{app7}) and moving to $a$-notation, we obtain 
\be
\hat\delta'' + \left( 2 - \frac{\Omega_M(a,z)}{2} \right)\frac{\hat\delta'}{a}-\frac{3}{2}\Omega_M(a,z)\frac{\hat\delta}{a^2} -
	\frac{2}{1+\hat\delta}\hat\delta'^2 - \frac{3}{2}\Omega_M(a,z)\frac{\hat\delta^2}{a^2} = 0.
\label{app8_ClassI}
\ee
It is worth noting again that here $a$ is a function of $t$ and $z$ while it was a function $t$ only in Class-II. 

Next, converting as before to notation using $\hat{G} \equiv \hat{\delta}/a$ for integration, we obtain
\be
\hat{G}'' + \left( 4-\frac{\Omega_M(a,z)}{2} \right)\frac{\hat{G}'}{a} + 2(1-\Omega_M(a,z))\frac{\hat{G}}{a^2} - \frac{2}{a}\frac{(\hat{G}+a\hat{G}')^2}{1+a\hat{G}}
	- \frac{3}{2}\Omega_M(a,z)\frac{\hat{G}^2}{a} = 0.
\label{app9_ClassI}
\ee
Now, we can express $\Omega_M(a,z)$ explicitly in terms of the scale factor, $a$, and the matter density parameter today, $\Omega_M^0(z)$ where a super- or subscript naught means that the parameter is evaluated today. One obtains 
\be
\Omega_M(a,z) = \frac{\Omega_M^0(z)}{\Omega_M^0(z)+a(1-\Omega_M^0(z))},
\label{app14_ClassI}
\ee
which we substitute into (\ref{app9_ClassI}) to finally get
{\small
\be
\hat{G}'' + \left(4-\frac{\Omega_M^0(z)}{2[\Omega_M^0(z)+a(1-\Omega_M^0(z))]} \right)\frac{\hat{G}'}{a}+2\left( 1-\frac{\Omega_M^0(z)}{\Omega_M^0(z)+a(1-\Omega_M^0(z))}\right)\frac{\hat{G}}{a^2}
	-\frac{2}{a}\frac{(\hat{G}+a\hat{G}')^2}{1+a\hat{G}} - \frac{3}{2}\left(\frac{\Omega_M^0(z)}{\Omega_M^0(z)+a(1-\Omega_M^0(z))}\right)\frac{\hat{G}^2}{a} = 0.
\label{app15_ClassI}
\ee
}

We first note that unlike the case of Class-II, here $\Omega_M^0(z)$ has a $z$-dependance and in order to perform the plots we will use it for a fixed value of $z$. Just as we specified for the quasi-local average density, we use this quasi-local average density number within a domain limited by such a constant value of $z$. 

Our results for a fixed value of $z$ are then identical in form to those of Class-II and are plotted and discussed in figures 1 and 2. The comments and remarks made for Class-II apply here as well.  

In summary, for both Class-I or Class-II, the Szekeres growth is found to be up to 5 times stronger than that of the linearly perturbed curved FLRW. Also, our plots in figure \ref{fig:density} for the density as a function of the scale factor for all classes and the cases that we explored show that the energy density diverges toward the initial singularity, as it should, and afterward it decreases monotonically with no other divergences during the matter-dominated era that we considered in this analysis. In order to explore further the strong growth found in the Szekeres models, we examine in the next section the shear and gravitational tides in the models.  

%%%%%%%%%%%%%%%%%%%%%%%%%%%%%%%%%%%%%%%%%%%%%%%%%%%%%%%%%%%%%%%%%%%%%%%%%%%
\section{The shear and the  tidal gravitational field in Szekeres models}
%%%%%%%%%%%%%%%%%%%%%%%%%%%%%%%%%%%%%%%%%%%%%%%%%%%%%%%%%%%%%%%%%%%%%%%%%%%%
%
In this section we analyze the time evolution of further physical quantities that can affect directly or indirectly the gravitational clustering and the growth rate of structure in the Szekeres models. The source being irrotational dust (i.e. zero vorticity and 4-acceleration), the next quantities of interest are the rate of shear and the tidal gravitational field. 
For this, we consider their corresponding scalar invariants.  

The discussion in this section applies to Class-I and Class-II since the evolution equations and scalar invariants are common to both classes. The distinction between the two treatments is that when we talk about Class-II, we use $\delta(t,x,y,z)$, $G\equiv\delta/a$, $M$, $a(t)$, and $\Omega_M(a)$, while we use for Class-I $\hat{\delta}(t,x,y,z)$, $\hat{G}\equiv\hat{\delta}/a$, $M(z)$, $a(t,z)$, and $\Omega_M(a,z)$ evaluated at a fixed value of $z$, as defined in the two previous sections. For simplicity of notation, we use here the formalism as used for Class-II but with observations and conclusions applying as well to Class-I with $z$ fixed. 

First we consider a shear scalar that is equal to twice the commonly used magnitude of the rate of shear \cite{Ellis1971,EllisVanElst1998} and defined as
\be
\sigma^2 = \sigma^a{}_b\sigma^b{}_a,
\ee
where the shear tensor components are given in equation (\ref{shear}) (note there is no factor $1/2$ in the definition). Computing this and taking the square root, we find
\be
\sigma = \sqrt{\frac{2}{3}}\frac{\dot{H}}{H}.
\label{sigma_a}
\ee
In order to analyze the time evolution of this scalar, we express it as a function of the scale factor and current measurable cosmological parameters. We can relate it to the density contrast via equation (\ref{identification}) and by noting that due to the  metric function dependencies, we have $\dot{H}=-\dot{F}$. Differentiating $\delta$ with respect to time allows us make the identification
\be
\frac{\dot{H}}{H}=-\frac{\dot{\delta}}{1+\delta}.
\label{sigma_b}
\ee
Now, using equation (\ref{app3}) and converting under/overdensities ($\delta$s) to growth rates ($G$s) as before, we obtain the shear in terms of the scale factor and the Hubble parameter $\mathbb{H}$:
\be
|\sigma| = \sqrt{\frac{2}{3}}\frac{G+aG'}{1+aG}a\mathbb{H}.
\ee
We use equation (\ref{app13}) to rewrite the Hubble parameter $\mathbb{H}$ in terms of its measured value today, $\mathbb{H}_0$, and that of the matter density today, $\Omega_M^0$:
\be
|\sigma| = \sqrt{\frac{2}{3}}\frac{G+aG'}{1+aG}\mathbb{H}_0\sqrt{\frac{\Omega_M^0}{a}-(
\Omega_M^0-1)}.
\label{ShearAbsValue}
\ee
\begin{figure}
\begin{center}
\begin{tabular}{|c|c|}
\hline
{\includegraphics[width=2.4in,height=2.in,angle=-0]{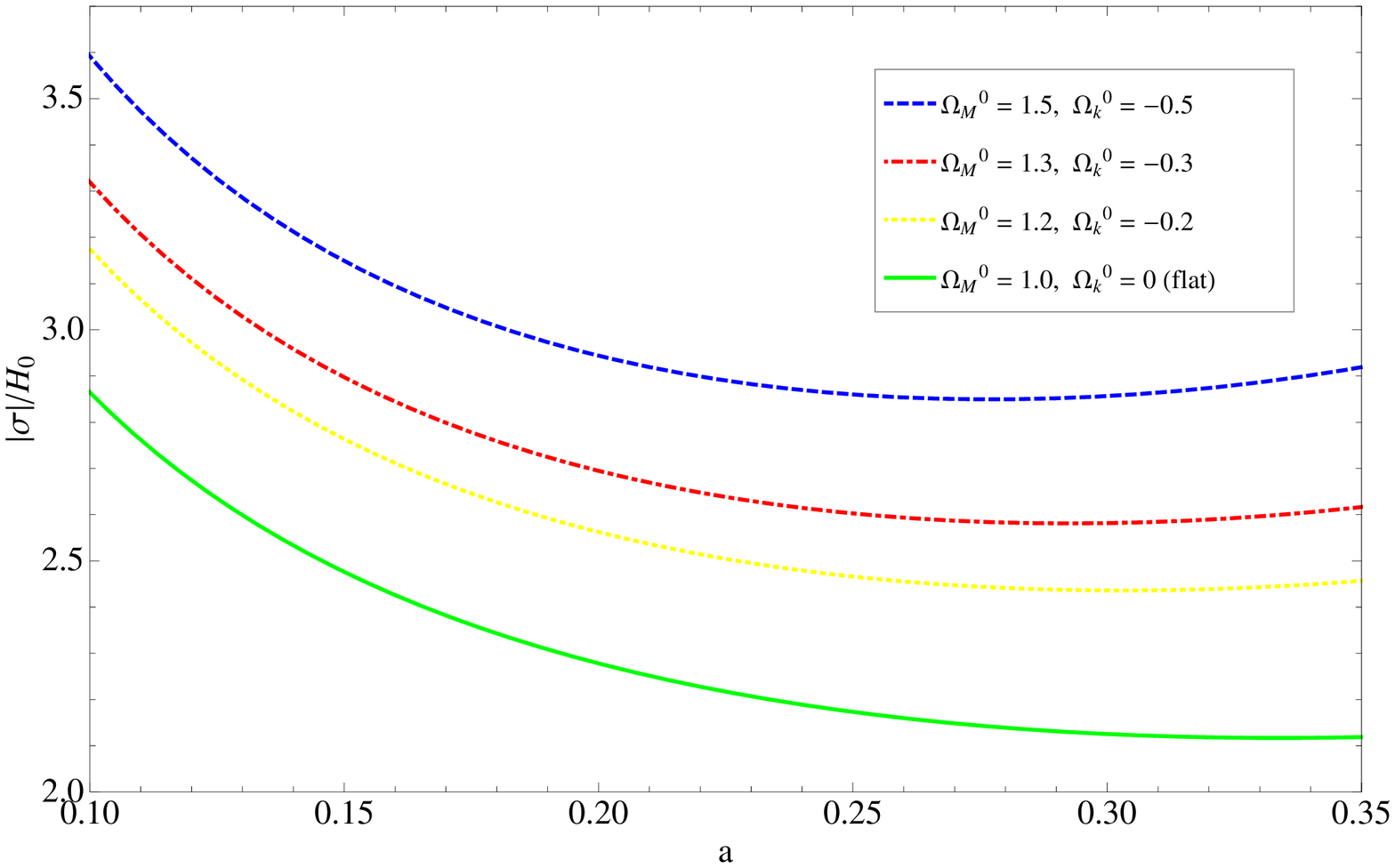}}&
{\includegraphics[width=2.4in,height=2.in,angle=-0]{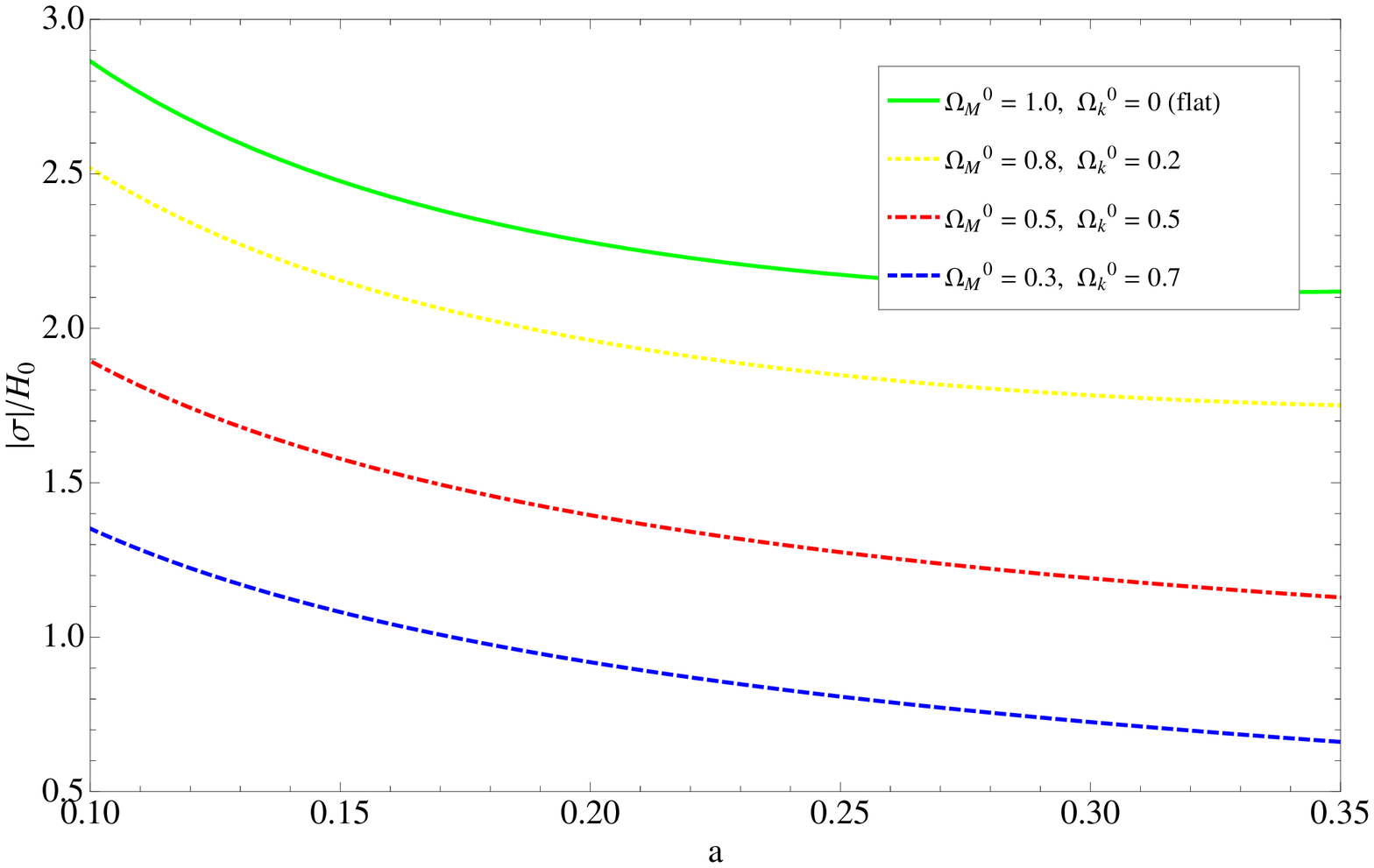}} \\
\hline
{\includegraphics[width=2.4in,height=2.in,angle=-0]{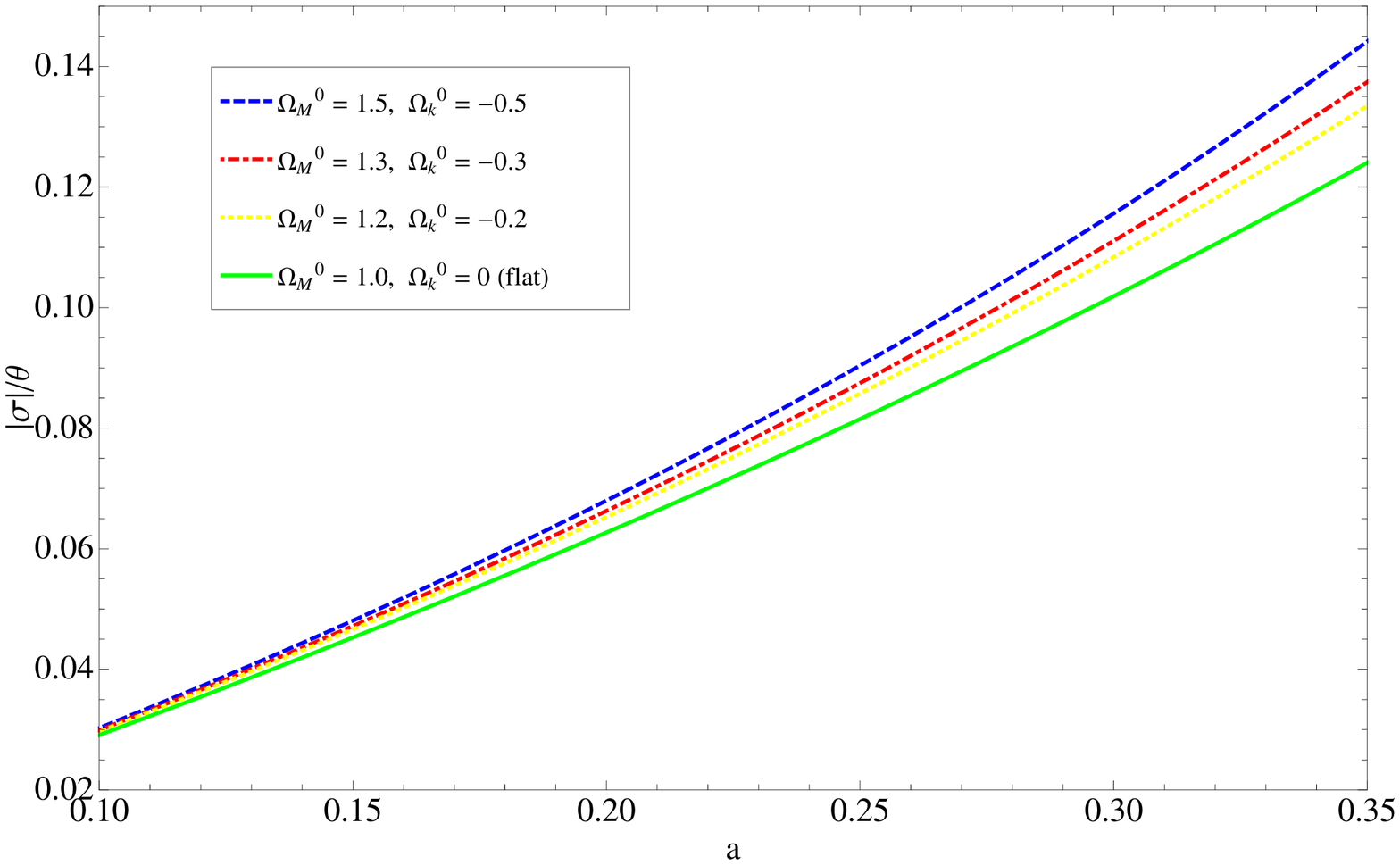}}&
{\includegraphics[width=2.4in,height=2.in,angle=-0]{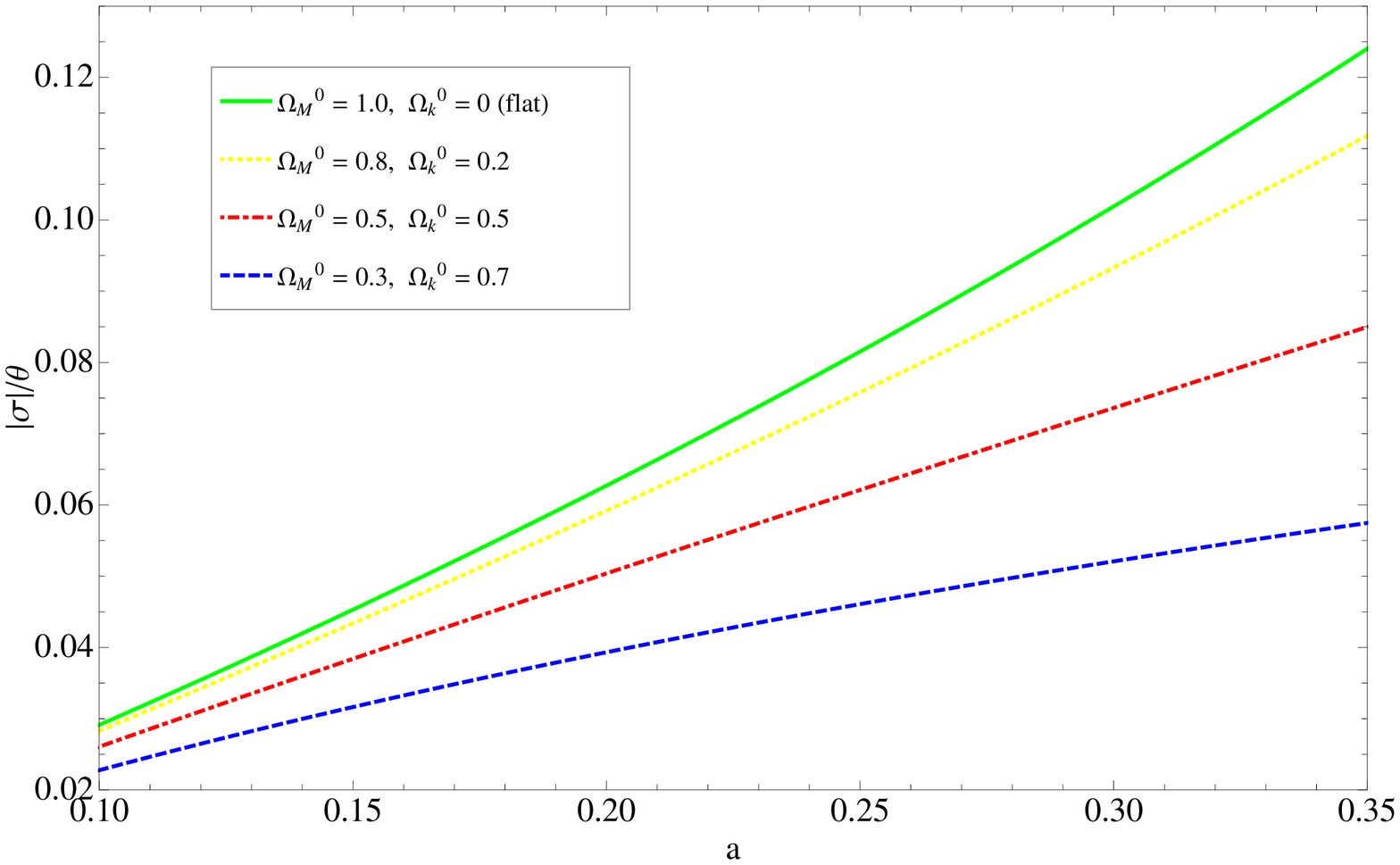}} \\
\hline  
\end{tabular}
\caption{\label{fig:ShearInSzekeres}
LEFT: Shear in positively curved Szekeres Class-II models (or Class-I models with a fixed value of $z$). The shear scalar $|\sigma|$ and scalar ratio $|\sigma|/\Theta$ are plotted as {functions} of the scale factor {well within the matter-dominated} cosmological era ($a\le 0.35$) (i.e., well before $a\approx 0.60${, the scale factor associated with matter and cosmological constant equality} in the FLRW-LCDM standard model \cite{IshakReport,Carroll}). The top {figure} shows that the shear {diverges when $a$ tends} to the initial singularity, while the bottom {figure} shows that the shear diverges less rapidly than the expansion scalar when approaching this initial singularity. As we discuss in {section V}, this presence of shear contributes to the enhancement of gravitational collapse and growth rate of large-scale structure in the Szekeres models. The amount of matter density (curvature) is varied as shown in the {plots, and} the spatially flat case is given for reference.
RIGHT: Shear in negatively curved Szekeres Class-II models (or Class-I models with a fixed value of $z$). The same observations {as for the positive case} hold here. All {four} figures here also show that the amplitude of the shear increases as the amount of matter density {does, so} the shear is stronger in the positively curved models {than in the flat case, which in turn is stronger than} for the negatively curved cases. This is consistent with the respective strengths of the growth rate as plotted in figure \ref{fig:Growth} for the flat, positively, and negatively curved cases. 
}
\end{center}
\end{figure}

Along with $|\sigma|$, another related quantity that is worth analyzing in our models is the shear scalar over the expansion scalar, $\frac{|\sigma|}{\Theta}$. This is particularly useful when looking at the early time divergences as we show further below. Using equation (\ref{expansion}) and  after a few steps we find that 
\bea
\Theta&=&\mathbb{H}\left[3-\frac{(G+aG')a}{1+aG}\right]\nonumber \\
      &=&\frac{\mathbb{H}_0}{a}\sqrt{\frac{\Omega_M^0}{a}-(\Omega_M^0-1)}\left[3-\frac{(G+aG')a}{1+aG}\right],
\eea
so that 
\be
\frac{|\sigma|}{\Theta}=\sqrt{\frac{2}{3}}\,\,\frac{a(G+aG')}{3+a(2G-aG')}. 
\label{ShearOverTheta}
\ee

Before plotting the time evolution of these scalars, we look again at the Raychaudhuri equation (\ref{Ray2}) where one can view the shear scalar, $\sigma^2$, as an ``effective source" that contributes along with the energy density, $\rho$, to cause stronger gravitational attraction and collapse. The equations (\ref{sigma_a}) and (\ref{sigma_b}) for $\sigma^2$ above show that indeed the shear introduces non-linear terms into the differential equations governing the growth. This causes the enhancement of the growth of large-scale structure seen for the Szekeres models examined in the previous section. The growth was found not only to be stronger than that of the homogeneous Einstein-de Sitter model, but also stronger than that of the spherical (inhomogeneous) growth, making clear the contribution of the shear in the Szekeres models.

Our plots for $|\sigma|$ and $\frac{|\sigma|}{\Theta}$ as given by equations (\ref{ShearAbsValue}) and (\ref{ShearOverTheta}) are shown in figure \ref{fig:ShearInSzekeres} for the flat and curved cases of the Szekeres Class-II models (the plots for Class-I with a fixed value of $z$ are exactly the same, and so are the conclusions drawn). The results show the time evolution {well within the matter-dominated} cosmological era and {well after the radiation-dominated} era. We chose $a\le 0.35${, which is well before $a\approx 0.60$, the scale factor associated with equality} between matter dominance and cosmological constant dominance in the FLRW-LCDM standard model \cite{IshakReport,Carroll}. In other words, we are here only concerned with the cosmological era fully dominated by matter (dust). The {upper plots show} that the shear approaches {infinity when $a$ tends to zero (the initial singularity),} while the {lower plots show} that the shear scalar diverges less rapidly than the expansion scalar when approaching this initial {singularity, since} the ratio is not diverging in that limit. These plotted behaviors for the exact scalars when approaching the initial singularity {are} consistent with the treatment of the asymptotic behaviors of the models derived by \cite{GoodeAndWainwright1982b} using other considerations. The {plots in the figure also show} that the amplitude of the shear increases as the amount of matter density {increases} so that the shear is stronger in positively curved models {than in} the flat case, which in turn is stronger than the shear for the negatively curved Szekeres models. This is consistent with the respective strengths of the growth rate of large-scale structure as shown in figure \ref{fig:Growth} for the flat, positively curved, and negatively curved cases in the previous section. 

Again, it is worth noting that our first goal here is to represent the growth during the {strictly matter-dominated} era with $a\le 0.35$. As shown in the plots, {no curves experience divergences after the initial singularity, and all} are far from any possible later time pancake singularity in the models \cite{GoodeAndWainwright1982b}. This is similar to the plots of the energy {density, which are certainly finite} during the {matter-dominated} era of interest (see figure \ref{fig:density}). Additionally, such pancake {singularities} can be avoided in some cases by the addition of a cosmological constant to this class of models \cite{MeuresAndBruni}.

For further exploration, it is informative to consider the next propagation equation of interest (after the Raychaudhuri equation), and that is for the shear, given by (e.g., \cite{Ellis1971,EllisVanElst1998}):
\be
\dot{\sigma}^a{\,}_b+\sigma^a{\,}_c \sigma^c{\,}_b+\frac{2}{3}\Theta{\,}\sigma^a{\,}_b-\frac{1}{3}(\delta^a{\,}_b+u^a\,u_b)  \sigma^2=-E^a{\,}_b,
\label{ShearPropag}
\ee 
where $E^a{\,}_b$ is the mixed tensor corresponding the {\textit {electric part}} of the Weyl conformal curvature tensor. The covariant tensor $E_{ab}$ is defined as (\cite{Ellis1971,EllisVanElst1998})    
\be
E_{ab} = C_{acbd}\,u^c\,u^d,
\label{Etensor}
\ee
where the Weyl tensor is given by    
\be
C_{abcd}=R_{abcd}+(g_{ad}R_{cb}+g_{bc}R_{ad}-g_{ac}R_{bd}-g_{bd}R_{ac})/2+R\,(g_{ac}g_{bd}-g_{ad}g_{bc})/6.
\label{eq:Weyl}
\ee

\begin{figure}
\begin{center}
\begin{tabular}{|c|c|}
\hline
{\includegraphics[width=2.3in,height=1.9in,angle=-0]{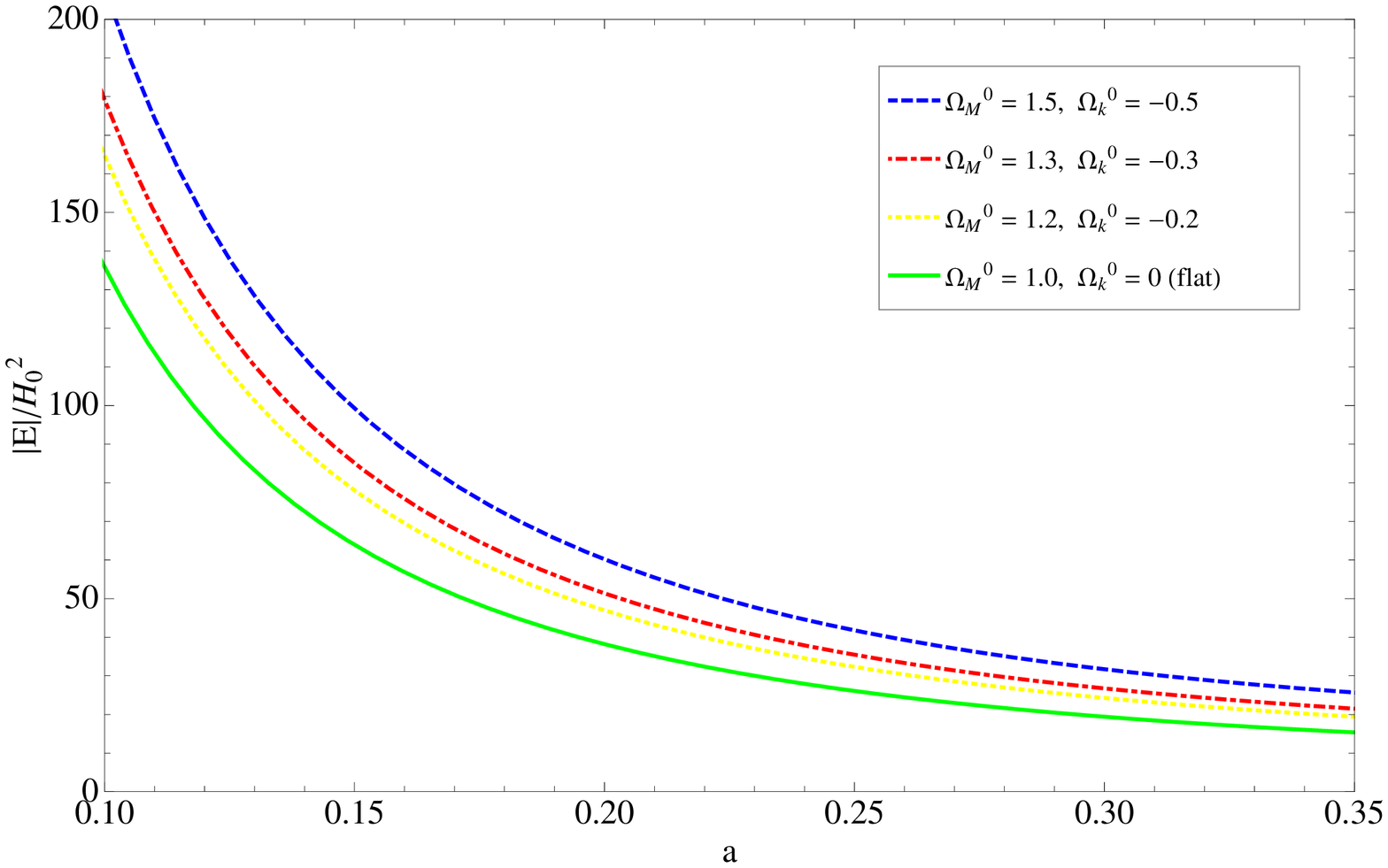}}&
{\includegraphics[width=2.3in,height=1.9in,angle=-0]{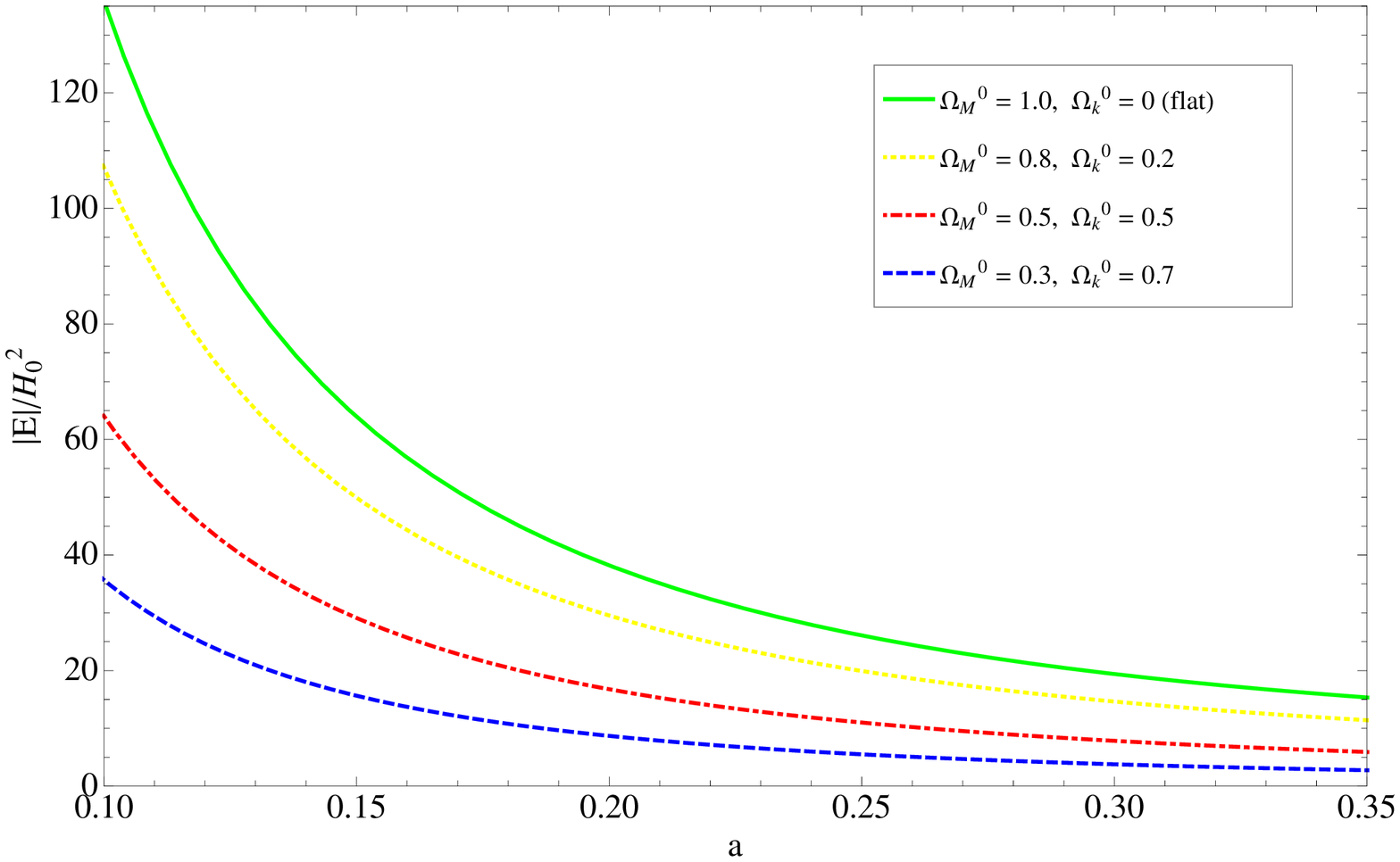}} \\
\hline
{\includegraphics[width=2.3in,height=1.9in,angle=-0]{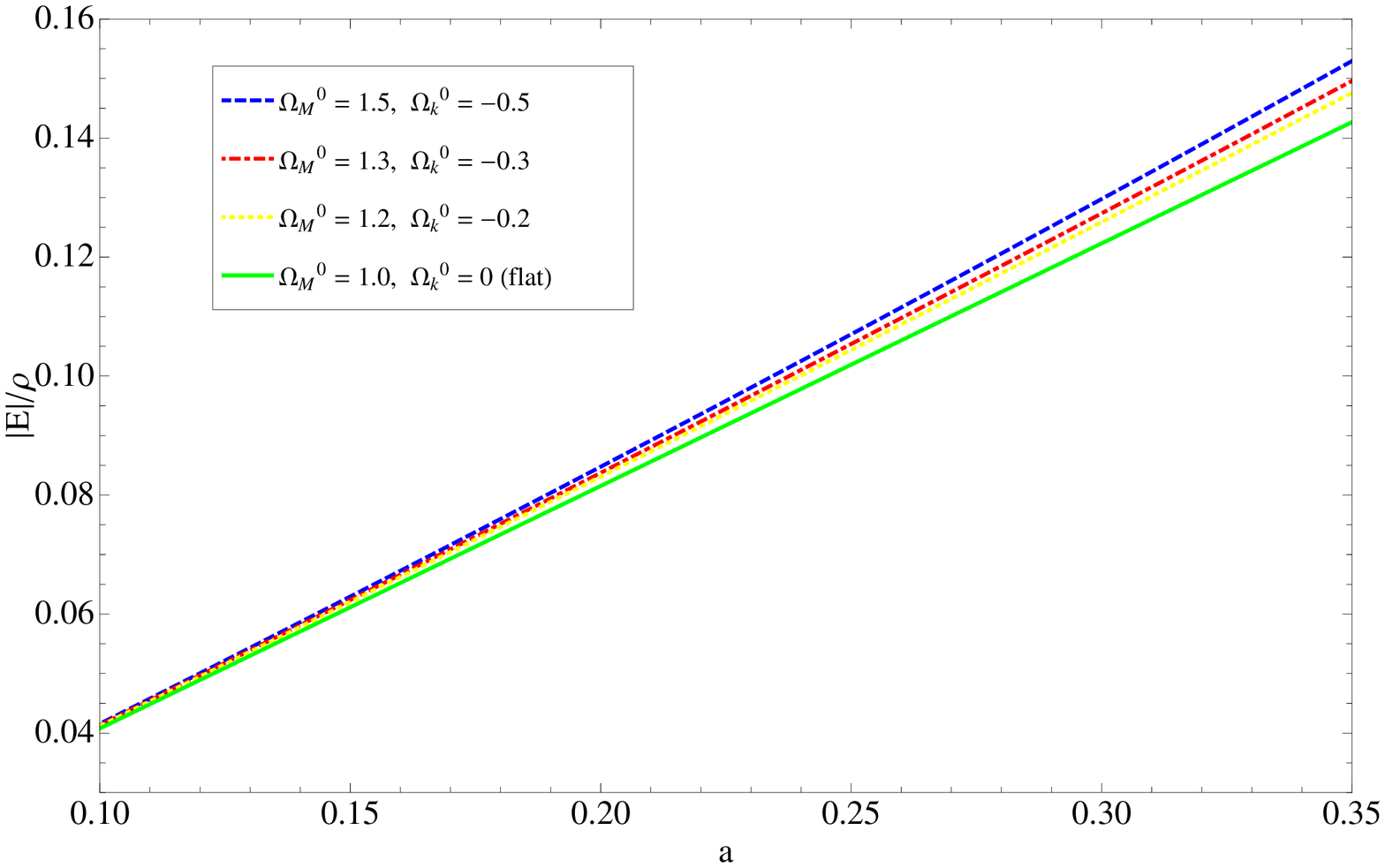}}&
{\includegraphics[width=2.3in,height=1.9in,angle=-0]{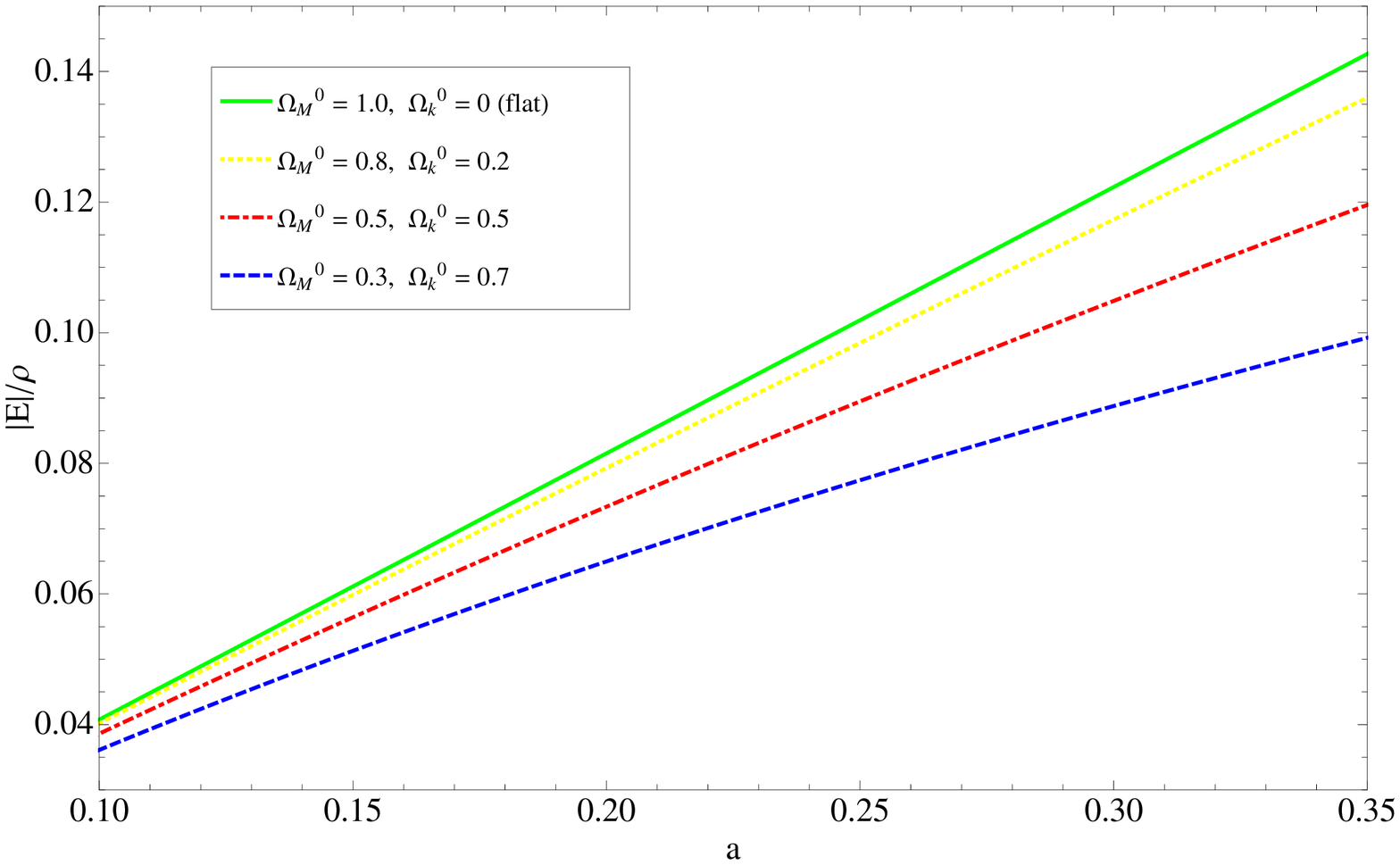}} \\
\hline  
\end{tabular}
\caption{\label{fig:WeylE}
LEFT: Tidal gravitational field in positively curved Szekeres Class-II models (or Class-I models with a fixed value of $z$). The scalar $|E|$ and scalar ratio $|E|/\rho$ are plotted as a function of the scale factor {well within the matter-dominated} cosmological era ($a\le 0.35$) (i.e., well before $a\approx 0.60${, the scale factor associated with matter and cosmological constant equality} in the FLRW-LCDM standard model \cite{IshakReport,Carroll}). The plots show that the time evolution of the tidal gravitational field is indeed very similar to that of the {shear, confirming} the statements made in section V about the tidal field inducing shear in the matter flow. Similarly to the shear plots, the {upper plots} show that the tidal field approaches {infinity} when $a$ {tends to zero, while the lower plots show} that the tidal field diverges less rapidly than the matter density when approaching this initial singularity. The amount of matter density (curvature) is varied as shown in the plots and the spatially flat case is given for reference.
RIGHT: Tidal gravitational field in {negatively} curved Szekeres Class-II models. The same observations {as for the positive case} hold here. 
All the figures here also show that the amplitude of tidal gravitational field  increases as the amount of matter density {does, consistent} with the amplitudes and evolution of the shear and the growth rate shown on previous plots. 
}
\end{center}
\end{figure}
As discussed in \cite{Ellis1971,EllisVanElst1998}, $E_{ab}$ represents the tidal gravitational field, and its presence in the shear propagation equation (\ref{ShearPropag}) shows how the tidal field induces shearing in the source flow. The shear then affects the gravitational collapse and clustering via the Raychaudhuri equation. Finally, it is well-known that in the Szekeres models the {\textit {magnetic part}} of the Weyl tensor, defined as $H_{ab}={}^{*}C_{acbd}\,u^c\,u^d$, vanishes \cite{Szekeres1,GoodeAndWainwright1982a}. (Here ${}^{*}C_{acbd}$ is the dual of Weyl tensor \cite{Ellis1971,EllisVanElst1998}).

Now, in order to analyze the contribution of the tidal gravitational field, we plot for the flat and curved Szekeres Class-II models the time evolution of the invariant  
\be
E=\sqrt{E^a\,_bE^b\,_a}.
\ee
For that, we first express the invariant in terms of the scale factor and measurable cosmological parameters as we did for the shear. We calculate the mixed components $E^a\,_b$ using equation (\ref{ShearPropag}) instead of the definition (\ref{Etensor}), leading to the simpler expressions 
\be
E^1\,_1=E^2\,_2=-2E^3\,_3=-\frac{1}{3}\frac{\ddot{H}}{H}+\frac{2}{3}\frac{\dot{H}}{H}\frac{\dot{a}}{a}.
\ee
It follows that 
\be
E=\pm \sqrt{\frac{2}{3}} \left(\frac{\ddot{H}}{H}+2\frac{\dot{H}}{H}\frac{\dot{a}}{a}\right).
\ee
Following the same steps as in the previous calculations for $\delta$, $\sigma$, and $\Theta$, we find
\be
E = \mp \sqrt{6}\frac{M}{a^3}\delta.
\label{weylE1}
\ee
The definition of $\Omega_M$ allows us to write
\be
\frac{M}{a^3}=\frac{\Omega_M}{2}\mathbb{H}^2.
\ee
We can then get $E$ in terms of $G$, $\mathbb{H}_0$, and $\Omega_M^0$ using equations (\ref{app13}) and (\ref{app14}). Equation (\ref{weylE1}) gives finally
\be
|E|=\frac{\sqrt{6}}{2}\frac{\Omega_M^0}{a^2}\mathbb{H}_0^2 G.
\ee
Similar to the shear over the expansion ratio, it is useful to define here the scalar ratio $\frac{|E|}{\rho}$ (see, for example, \cite{GoodeAndWainwright1982b}) in order to study early time divergences. Using the expression (\ref{RhoToPlot}) for the energy density we can write this ratio as
\be
\frac{|E|}{\rho}=\frac{1}{\sqrt{6}}\frac{aG}{1+aG}.
\ee
Our plots for the tidal gravitational field scalar $|E|$ and the ratio $\frac{|E|}{\rho}$ are given in figure \ref{fig:WeylE} for Class-II (the plots for Class-I with a fixed value of $z$ are exactly the same, and so are the conclusions drawn). The scalar $|E|$ and scalar ratio $|E|/\rho$ are plotted as a function of the scale factor {well within the matter-dominated} cosmological era. We find that the time evolution of the tidal field is very similar to that of the shear and confirms the discussion above about the tidal field inducing shear (according to the shear evolution equation (\ref{ShearPropag})) in the source matter flow. We also see the same behavior as for the shear at earlier times where the tidal field {diverges as $a$ approaches the initial singularity, but} the ratio $\frac{|E|}{\rho}$ does not, because the tidal field diverges less rapidly than the matter density {as the scale factor goes to zero}. The amount of matter density (curvature) is varied, and the figures show how the amplitude of tidal gravitational field increases as the amount of matter density does, consistent with what we observed for the shear and the growth rate plots. 

%%%%%%%%%%%%%%%%%%%%%%%%%%%%%%%%%%%%%%%%%%%%%%%%%%%%%%%%%%%%%%%%%%%%%%
\section{Concluding remarks}
%%%%%%%%%%%%%%%%%%%%%%%%%%%%%%%%%%%%%%%%%%%%%%%%%%%%%%%%%%%%%%%%%%%%%%
%%%%
%
We considered the growth rate of large-scale structure in the Szekeres inhomogeneous cosmological models written in the Goode and Wainwright representation. Using the Raychaudhuri equation, we derived exact equations for the growth rate for the two Szekeres classes. We explicitly expressed the equations in terms of the under/overdensity and measurable cosmological {parameters, leading} to further insights on the growth rate of structures in the Szekeres models and also putting them in a framework close to comparison with cosmological observations. 

In Class-II, the background density is only a function of time, so we defined the under/overdensity in the standard way, while in Class-I we defined an invariant under/overdensity using the quasi-local averaged density over a spatial domain. For both classes, we found that writing these equations in terms of these under/overdensities instead of the metric functions allows for the growth equations to be remarkably split into two meaningful parts. The first is identical to the usual growth equations of a perturbed matter-dominated FLRW model. The second part is similar to second-order perturbations, but here the equations derived are all exact, and this part {represents} the nonlinearity of the Szekeres exact solution. 

We integrated numerically the exact equations obtained for flat Class-II, curved Class-II, and curved Class-I models. We note that flat Class-I models have no growing modes and so are of no interest to our purpose here. For curved Class-I models, we use a domain delimited by a fixed value of $z$. In all the cases, we found that the Szekeres growth rate is up to 3-5 times stronger than that of the perturbed FLRW models. The Szekeres growth (with a corresponding Einstein-de Sitter background) is also found to be stronger than the well-known nonlinear spherical collapse with the difference between the two increasing with time. This shows that the growth is stronger and different when we use the more general Szekeres inhomogeneous models where shear and tidal gravitational fields are present in the matter source.  

In order to explore this Szekeres strong growth further, we derived explicit expressions for the shear scalar and tidal gravitational field part of the Weyl tensor, again all in terms of the scale factor and measurable cosmological parameters. Our analysis shows and confirms how the shear acts in the Raychaudhuri equation like an ``effective source" along with the energy density to enhance gravitational collapse and produce stronger growth of structure in the Szekeres models. We also derived and plotted the time evolution of tidal gravitational field which induces shearing in the cosmic fluid flow, as can be seen from the shear evolution equation.  

In this analysis, we focused our interest and results to be {well within the matter-dominated} cosmological era. That is {well after the radiation-dominated era and also well before the equality time} between matter dominance and cosmological constant dominance in the FLRW-LCDM standard model. In other words, we are here only concerned with the cosmic era fully dominated by matter. We also plotted the matter energy density during this era and found that it starts diverging {(as expected)} close the initial singularity, but after that it decreases monotonically with no other divergences during this interval, showing that we are far away from any possible later time pancake singularity in the models. Additionally, such pancake singularities can be avoided by the addition of a cosmological constant to the models as shown in other works. We also plotted the time evolution of other scalars, such as the shear over the expansion and the tidal gravitational field over the energy density, and found that they tend to zero when approaching the initial singularity, in agreement with previous results that used other considerations. 

It is worth noting that the enhancement of the growth found here in the
Szekeres models during the matter-dominated era could suggest a substitute 
to the argument that dark matter is needed when using an FLRW-LCDM model during the matter-dominated era in order to explain the enhanced growth and all the large-scale structure that we observe today. Indeed, it is a well-known argument in FLRW models that dark matter is necessary to enhance the growth during the matter-dominated era in order to explain all the presently observed galaxy clusters and superclusters. In the Szekeres models the enhanced growth seems present with no requirement of dark matter. A similar conclusion was reached {in, for example,} \cite{DodelsonEtAl} but from analyzing growth in a modified gravity model. Of course, inhomogeneous models with {the} presence of shear and a tidal gravitational field should also be explored to analyze galaxy rotation curves and other arguments in support of dark matter. We plan to explore this in follow-up work. 

Finally, the current results for the growth in Szekeres expressed within standard schemes of growth rate studies and in terms of observable cosmological parameters will be useful toward building a thorough framework where inhomogeneous Szekeres cosmological models can be compared to various current and future cosmological data sets. Comparison of the growth rate in Szekeres to available and future data probing the growth is out of the scope of the current paper but will need to be done in future and follow up works.   
%%
%
%%%%%%%%%%%%%%%%%%%%%%%%%%%%%%%%%%%%%%%%%%%%%%%%%%%%%%%%%%%%%%%%% 
\acknowledgements
%%%%%%%%%%%%%%%%%%%%%%%%%%%%%%%%%%%%%%%%%%%%%%%%%%%%%%%%%%%%%%%%%%
%
We thank R. Sussman for valuable comments and M. Troxel for proof-reading the manuscript. MI acknowledges that this material is based upon work supported in part by NASA under grant NNX09AJ55G, by Department of Energy (DOE) under grant DE-FG02-10ER41310, and that part of the calculations for this work have been performed on the Cosmology Computer Cluster funded by the Hoblitzelle Foundation.
%
%%%%%%%%%%%%%%%%%%%%%%%%%%%%%%%%%%%%%%%%%%%%%%%%%%%%%%%%%%%%%%%%%

%%

\begin{thebibliography}{}
\bibitem{Note}In references \cite{Sussman1,SussmanAIP,Sussman2,SussmanBolejko} the $\hat\delta$ that we defined here was noted as $\Delta^{(\rho)}$. Also, as discussed in \cite{Sussman2} the sign of $\Delta^{(\rho)}$ and the identification of under- overdensity in inhomogeneous cosmological models invoke some subtleties. These and other points should be explored within Szekeres models in future work.
\end{thebibliography}
\end{document}